\documentclass[preprint]{acmsiggraph}

\TOGonlineid{0160}
\TOGvolume{}
\TOGnumber{}
\TOGarticleDOI{}
\TOGprojectURL{}
\TOGvideoURL{}
\TOGdataURL{}
\TOGcodeURL{}
\usepackage{ dsfont }
\usepackage{rotating}
\usepackage{wrapfig}
\usepackage[table]{xcolor}
\definecolor{LightCyan}{rgb}{0.88,1,1}
\title{Voronoi Grid-Shell Structures}

\author{Nico Pietroni$^{1}$\thanks{e-mail: nico.pietroni@isti.cnr.it}
\and Davide Tonelli$^{2}$\thanks{e-mail: davide.tonelli@dic.unipi.it}
\and Enrico Puppo$^{3}$\thanks{e-mail: puppo@disi.unige.it}
\and Maurizio Froli$^{2}$\thanks{e-mail: m.froli@dic.unipi.it}
\and Roberto Scopigno$^{1}$\thanks{e-mail: roberto.scopigno@isti.cnr.it}
\and Paolo Cignoni$^{1}$\thanks{e-mail: paolo.cignoni@isti.cnr.it}
\\
$^1$ISTI, CNR, Italy $\quad$
$^2$University of Pisa $\quad$
$^2$University of Genova $\quad$
}

\pdfauthor{ Nico Pietroni,Davide Tonelli,Enrico Puppo,Maurizio Froli,Roberto Scopigno,Paolo Cignoni}


\begin{document}

 \teaser{
   \includegraphics[height=2.05in]{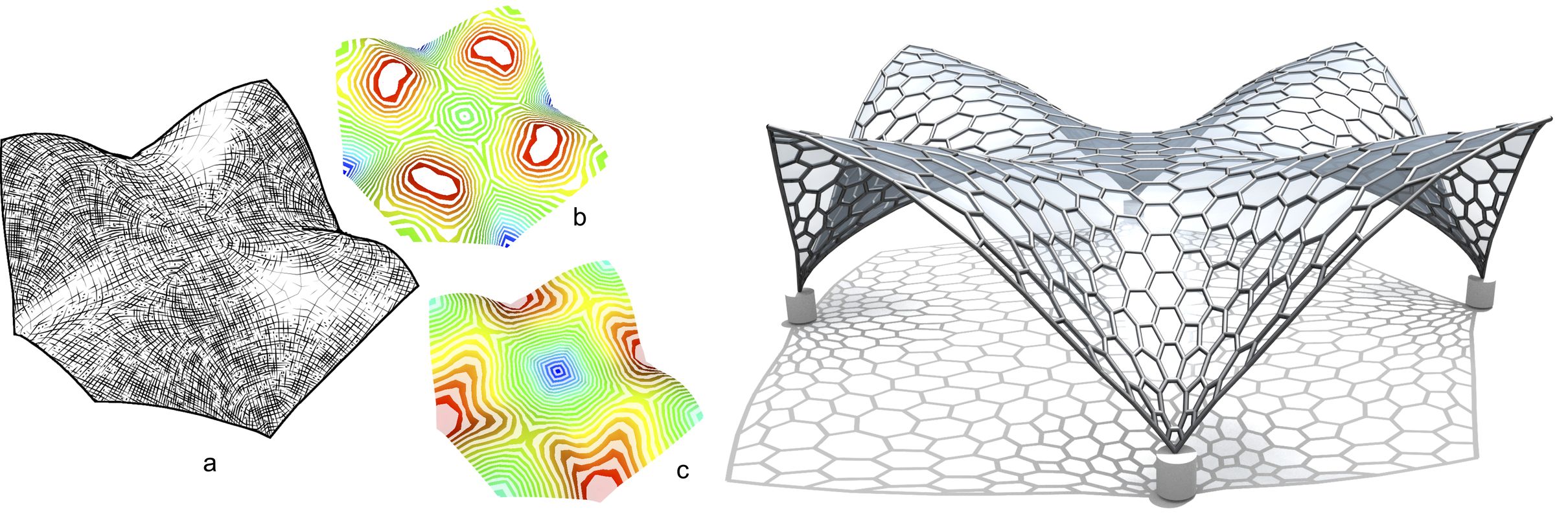}
   \caption{We perform a FEM static analysis of the input surface to obtain a stress tensor field, which is decomposed into a double orthogonal line field (a), an anisotropy scalar field (b) and a density scalar field (c). Then we build an Anisotropic Centroidal Voronoi Tessellation having its elements sized and aligned according to the stress tensor field; this tessellation is optimized for symmetry and regularity of faces. 
   The resulting grid-shell is hex-dominant and it is designed to fulfill the required static properties.}
 \label{fig:teaser}
 }

\maketitle

\begin{abstract}

We introduce a framework for the generation of grid-shell structures that is based on Voronoi diagrams and allows us to design tessellations that achieve 
excellent static performances. 
We start from an analysis of stress on the input surface and we use the resulting tensor field to induce an anisotropic non-Euclidean metric over it. 
Then we compute a Centroidal  Voronoi Tessellation under the same metric. 
The resulting mesh is hex-dominant and made of cells with a variable density, which depends on the amount of stress, and anisotropic shape, which depends on the direction of maximum stress. 
This mesh is further optimized taking into account symmetry and regularity of cells to improve aesthetics.
We demonstrate that our grid-shells achieve better static performances with respect to quad-based grid shells, while offering an innovative and aesthetically pleasing look.
\end{abstract}

\begin{CRcatlist}
  \CRcat{I.3.5}{Computer Graphics}{Computational geometry and object modeling}{Curve, surface, solid and object repres.}
\end{CRcatlist}

\keywordlist

\TOGlinkslist

\copyrightspace

% !TEX root = main.tex
\section{Introduction}\label{sec:intro}

Grid-shells, such as steel-glass structures, have been used for about forty years in architecture \cite{Otto:1995}.
While triangle-based grid-shells seem unbeatable from the point of view of strength, quad-based structures have become popular in the last decade, because of their improved aesthetics and nice mathematical properties.
%as well as their ability to align mesh elements to principal curvature directions or conjugate directions.
Conversely, there exist fewer studies on more general polygonal structures, most of which are focused on improving mesh geometry for a given topology \cite{Pottmann:2014}.
%Also, there exist few examples of grid-shell design driven by the surface statics \cite{Schiftner:2010}.
%\textbf{Davide puoi indicare fonti su progettazione di grid-shell guidata dalla statica? Anche se si riferiscono a cose fatte "a mano".}  

In this paper, we introduce a framework for the generation of grid-shell structures that is based on Anisotropic Centroidal Voronoi Tessellations. 
Our method is driven by the statics of the input surface and it is aimed at improving the strength of the grid-shell as well as its aesthetics. 
Voronoi diagrams appear in nature in many forms, and in several cases they are related to light and strong structures. 
For instance, bones have a Voronoi-like porous structure, with a higher concentration of material where the bone undergoes more stress.
This natural principle has also been applied recently to object design for 3D printing \cite{Lu:2014}.
We follow a similar approach to design our grid-shells, by concentrating more cells of smaller size in zones subject to higher stress, while aligning the elements of our grid to the maximum stress direction.

We start at an input surface and we aim at producing a grid-shell that approximates this surface closely. 
We first perform a static analysis of the surface, from which we obtain an anisotropic,  non-Euclidean metric described by the stress tensor.
Next we deform the surface, similarly to \cite{Panozzo:2014}, in order to transform this anisotropic metric into an Euclidean metric on the deformed surface.
We perform Poisson sampling on the deformed surface and we compute a Centroidal Voronoi Tessellation of sampled points. 
This diagram is mapped back to the original surface to obtain an Anisotropic Centroidal Voronoi Tessellation.

We can control the variation of density and anisotropy of our meshing through two simple parameters.
We apply geometric optimization to follow surface symmetries and to improve the local shape of faces of our mesh, making them closer to the faces of Archimedeal solids; this geometric optimization phase greatly contributes to improve the aesthetics of our grid-shells, and it also slightly improves its static performances.

We show that the structures generated with our approach, thanks to the great flexibility of Voronoi diagrams, are able to adapt well to the needs of architects and to the designed shapes, while achieving better static behaviour with respect to quad-based grid-shells.
%in some case even better than grid-shells based on isotropic triangle meshes. 

% !TEX root = main.tex
\section{Related Work}\label{sec:star}

\paragraph{Voronoi diagrams.}
The Voronoi Diagram (VD) \cite{Aurenhammer:1991} is a fundamental geometric data structure; within the scope of this work, we discuss only the literature about remeshing techniques based on Centroidal Voronoi Tessellation (CVT) \cite{Du:1999}.  
VD's can be defined over 3D surfaces using geodesic distance. 
This can be done in various ways, either using discrete approximation of geodesic distance \cite{Peyre:2006}, or using parametrization techniques to bring the problem onto a 2D domain \cite{Alliez:2005}. 
Valette and Chassery \cite{Valette:2004} compute an approximated CVT, based on a discrete global minimization approach.
There exist several proposals for computing an Anisotropic CVT (ACVT) under a Riemaniann metric \cite{Du:2005,Sun:2011,Valette:2008}.
Some recent efficient techniques are based on projection of the domain to a 6D space in which the metric becomes Euclidean \cite{Levy:2013,Zhong:2013}. 
Panozzo et al.\  \shortcite{Panozzo:2014} show that a simpler deformation of the surface in 3D is sufficient to get an approximated Euclidean metric, provided that the changes of scale and anisotropy induced by the original metric are not too high.

\paragraph{Architectural geometry.}
%\cite{Wallner:2011}
Most contributions in this field are concerned with the optimization of geometric properties of polygonal meshes approximating a free-form surface.
Many works address the planarity of faces, such the construction of PQ (planar quad) meshes \cite{Liu:2006,Liu:2011,Tang:2014,Schiftner:2010,Yang:2011,Zadravec:2010}, CP (circle packing) meshes \cite{Schiftner:2009}, and polygonal hex-dominant meshes \cite{Cutler:2007,Pottmann:2014,Schiftner:2009,Troche:2008}.
%,Wang:2008}.
Others try to build meshes from a restricted number of tiles or molds \cite{Eigensatz:2010,Fu:2010,Singh:2010,Zimmer:2012}.
A few works address the realization of support structures, parallel meshes and torsion-free meshes \cite{Pottmann:2007,Pottmann:2014,Tang:2014}.
Among these works, only few focus on the design of a grid topology \cite{Cutler:2007,Liu:2011,Schiftner:2010,%Troche:2008,
Zadravec:2010} and just Schiftner and Balzer \shortcite{Schiftner:2010} take into account statics.
Pottmann et al.\ \shortcite{Pottmann:2014} mention the possibility of building grid-shell structures from either CVT or ACVT to obtain hex-dominant meshes; they do not further investigate the underlying design principles, though.

\paragraph{Statics of grid-shell structures.}

Grid-shell structures are a modern response to the ancient need of covering long span spaces. 
%Basically they are the discrete version of a continuous shell structures, with beams along the edges and panels or openings in place of the faces.
%Grid-shells 
They are compressive structures, i.e. the principal stress comes mainly from axial forces, and this explains the deep interconnection between them and masonry structures. 
%Because of this, geometry plays a fundamental role in the design of grid-shells: a 
A robust as well as light grid-shell can be obtained only through a form-finding process, aimed at finding the \emph{funicular surface} (surface which stands under compression-only stresses) that fits the given boundary constraints \cite{BulKni,Ogawa,Otto:1995}. 
It is well known \cite{BulKni} that the form of quad meshes is maintained only if the joints are able to develop bending moments, while triangular meshes maintain their form even if the joints are hinges.

Thrust Network Analysis \cite{Block:2009}, a recent form-finding method derived from graphics statics, is specific to masonry. 
An extension of this method was recently introduced by Tang et al.\ \shortcite{Tang:2014}, which directly allows for grid-shell form finding:
%This latter method 
not only it computes the target funicular surface, but it also optimizes the positions of edges.
%(enrico aggiungi pure tutti i riferimenti che vuoi. se lo ritieni interessante possiamo aggiungere prima di block 2009 anche riferimenti ad altri metodi come il dynamic relaxation ed il force density).\\
In Section \ref{sec:results}, we compare some of our results with grid-shells obtained with this latter method.

The connectivity of the mesh %is very important since it 
is directly related to the  %position and distribution of the edges on the surface, and hence to the 
load bearing capacity of the grid-shell. 
While triangular meshes are more rigid and stronger than any other competitor, polygonal meshes 
%are worse in terms of both rigidity and strength, but they 
% allow for more light to seep through the envelope, and 
have some advantages in terms of ease of construction and lend themselves to the design of torsion-free structures.
%In Architectural Geometry several studies have been carried out to evaluate the geometric properties of quadrilateral meshes for fabrication purposes (e.g. planarity, single curvature, conical or circular property of the faces etc..) (citare Liu delle PQmeshes, Pottman un pò a caso..). 
%Simultaneously s
Some comparative parametric analyses \cite{Malek:2013} have been carried out about the influence of the remeshing pattern on the grid-shell load bearing capacity. 
There exist surprisingly few studies about the optimal (in a structural sense) connectivity and distribution of edges \cite{Schiftner:2010}, although probably these are -- in conjunction with the surface shape -- the most influential parameters that govern the structural behavior of the grid-shell. 
%Our contribution tackles exactly this topic: the effect of connectivity and distribution of edges on the structural perfrormance of the grid-shell.
%
%We've arranged a set of comparative structural analyses in order to first evaluate the structural performances of the specimen, and then to compare them to other \textit{equivalent} state of the art (triangular or quadrilateral) grid-shells. The 
For our comparative experiments of different grid-shells for a given shape, we adopt the equivalence criterion of simultaneous equal total mass and equal total length of edges, as in \cite{Malek:2013}.
%: infact we've experienced that the total mass equivalence cannot guarantee by itself the structural capability equivalence.\\
%We've decided to consider only the static properties of the grid-shell structures and among these we've selected just two comparison quantities:
%\begin{enumerate}
%\item \textit{collapse load factor} - from non linear static analysis;
%\item \textit{maximum displacement norm} - from linear static analysis.
%\end{enumerate}
%The collapse load factor is a scalar quantity which represents how many time the load condition can be increased without reaching the structure's collapse: it quantifies the overall \textit{strength} if the structure. The maximum displacement norm comes from a linear elastic analysis, and therefore portrays the overall \textit{elastic stiffness} of the structure. An ideal grid-shell should have an as high as possible load factor and an as low as possible maximum displacement norm.\\ We've chosen these quantities for comparison as they're scalar quantities (and therefore extremely simple to compare) and heavily representative of the structural behavior.

jpg% !TEX root = main.tex
\section{Surface Metric from Static Analysis} \label{sec:static}

%Our main objective is to use stress analysis techniques to drive the creation of an optimized grid-shell structure whose topology adapts to stresses and loads of an input surface. 
%Intuitively speaking, we want to drive the construction of an ACVT in which the density of the tessellation reflects the amount of stress of the structure and the Voronoi regions are aligned with the maximal stress direction.

The first step of our pipeline is to perform a linear static analysis of the input surface. % shape. 
More precisely we analyze a continuous shell subject to uniform projected load and with all boundary nodes pin-restrained. 
%Then we extract three data sets for each face: the direction cosines, the \textit{principal directions} (which are a cross field) and the \textit{principal stresses}.
This analysis returns a tensor field, whose eigenvectors and eigenvalues represent the principal directions and the principal stresses at each point, respectively.  
In structural mechanics, 
%the principal directions (three for each point of a 3D body - for a shell the third is always the normal to the face) are those with respect to which the stress tensor becomes diagonal - thus making shear stresses equal to 0, whereas the principal stresses are the diagonal terms of the stress tensor in the principal reference. Then the 
\emph{isostatic lines} are pairs of curves on the shell, which are always tangent to a principal direction, hence always orthogonal to each other. 
Concentrating the material along the isostatic lines is a good way to improve the structural performance of a structure:
%: in fact they locate the zones where there's the utmost need for strength and moreover most materials (e.g. masonry but steel too) show much higher normal stress strength than shear strength. That's 
in a nutshell, this is what we try to do with our contribution.

\begin{figure}[t]
  \centering
  \includegraphics[width=1in]{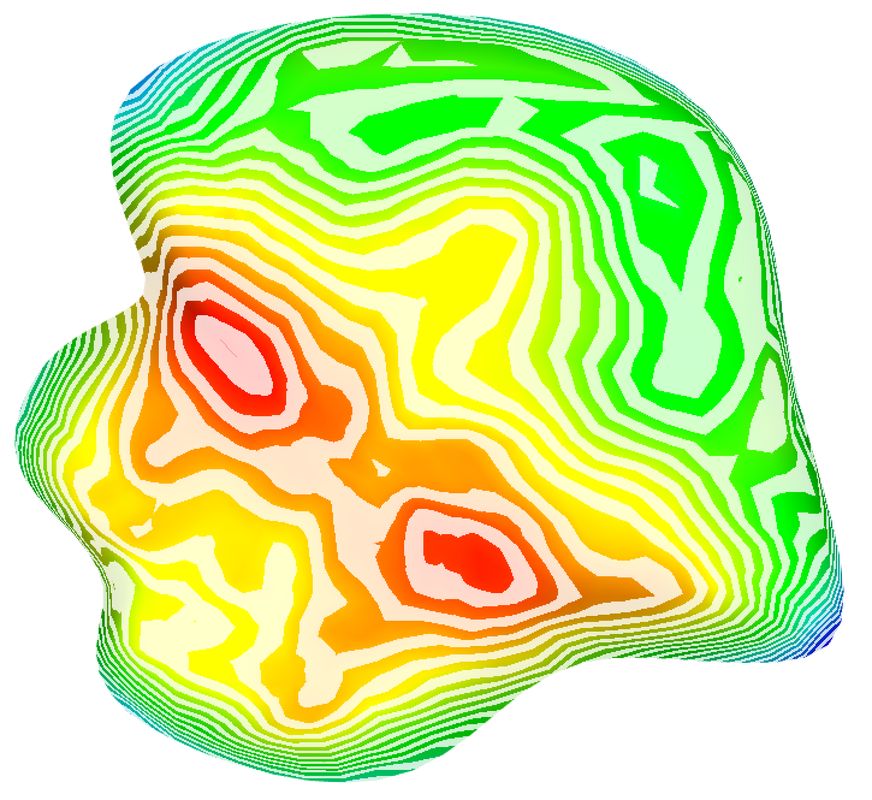}
  \includegraphics[width=1in]{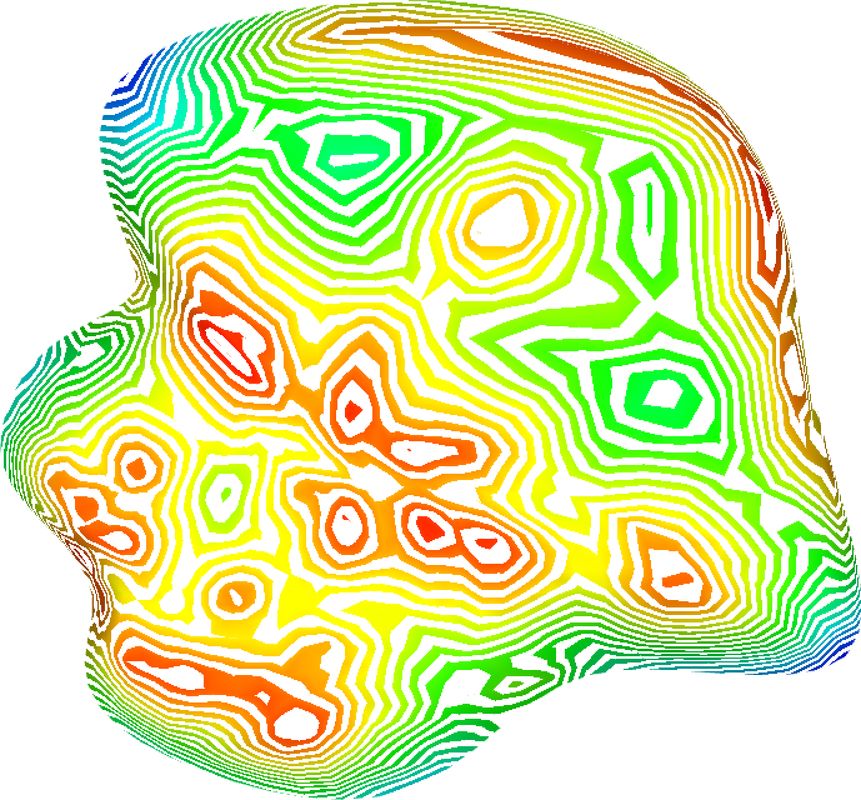}
  \includegraphics[width=1in]{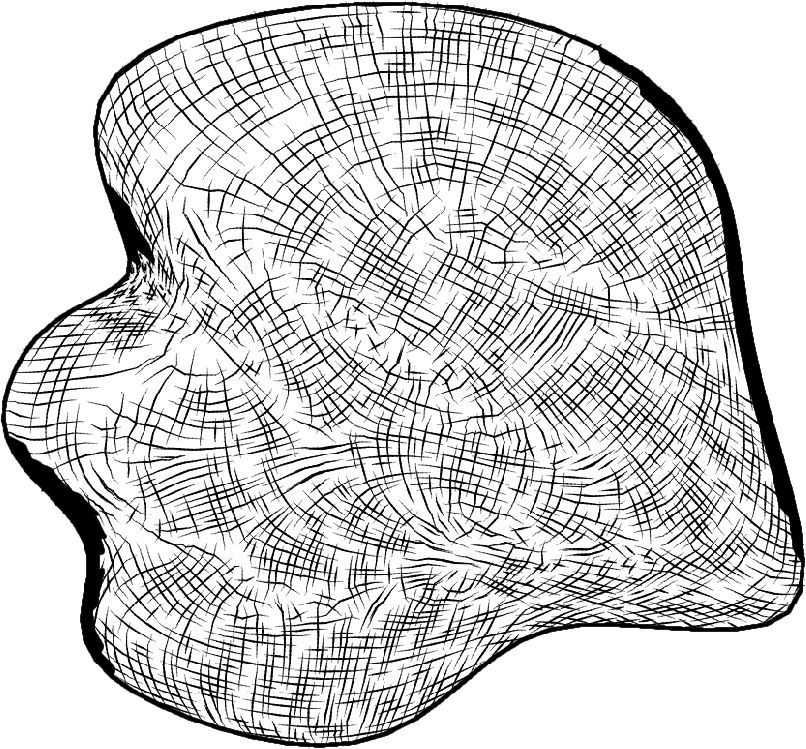}
  \includegraphics[width=1in]{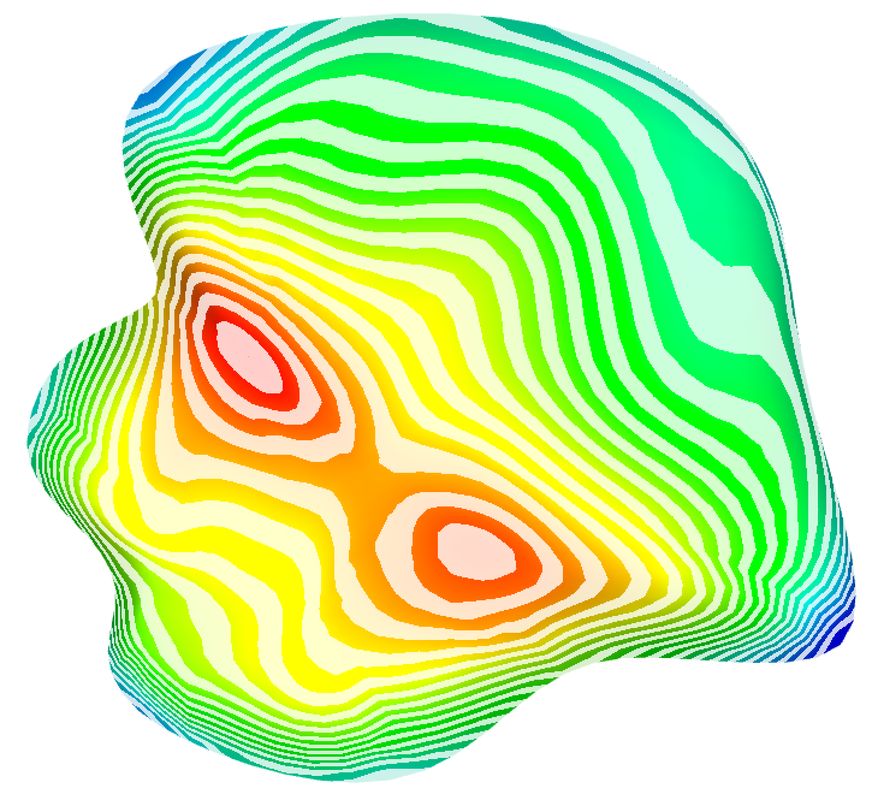}
  \includegraphics[width=1in]{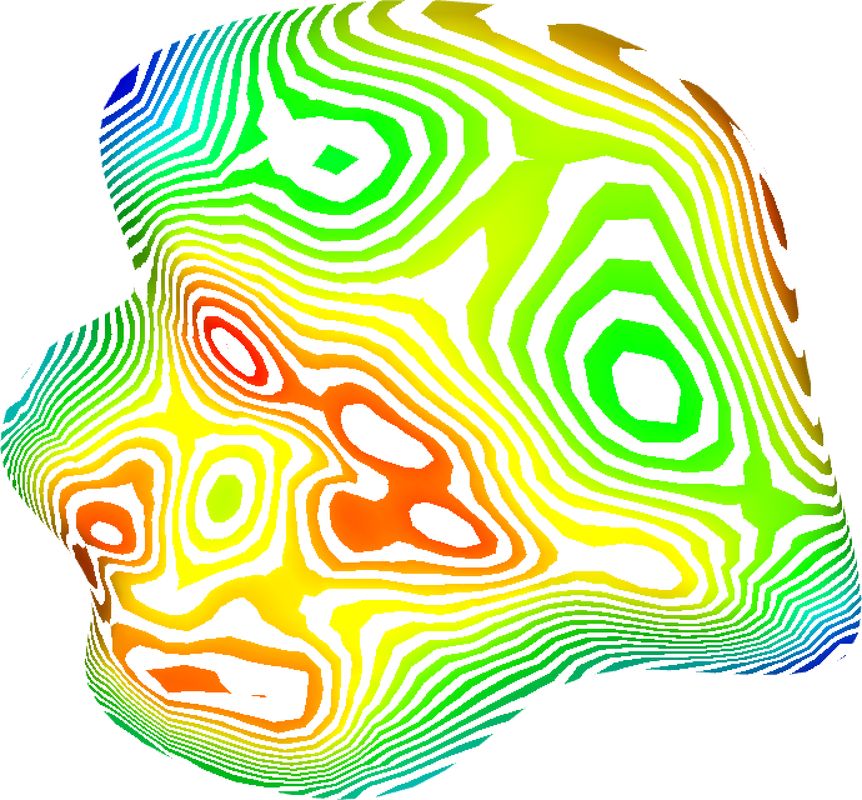}
  \includegraphics[width=1in]{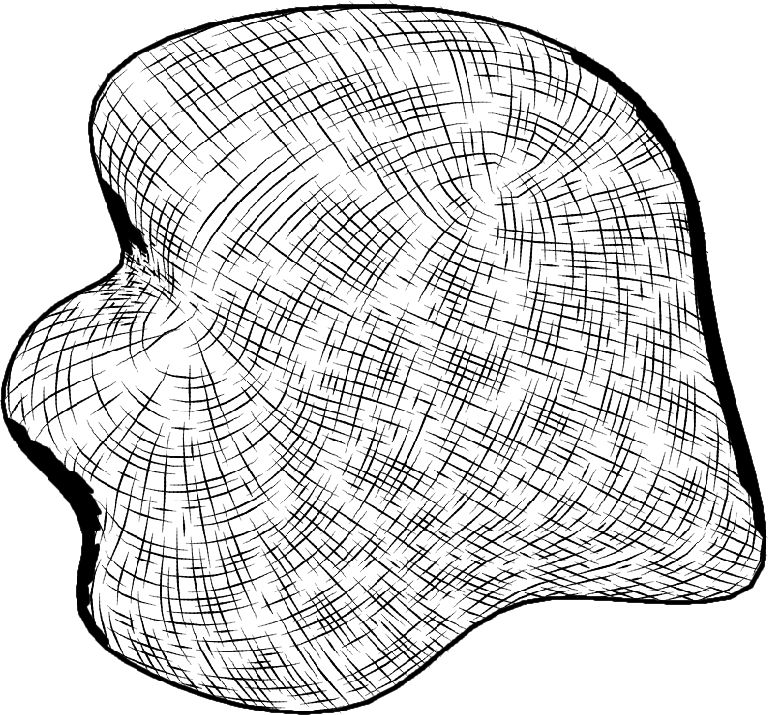}
  \caption{Smoothing and saturating the \emph{density} (left), the \emph{anisotropy} (middle), and the two orthogonal line fields (right) of the Botanic model: upper side original, lower side smoothed.}
  \label{fig:static_smoothing}
\end{figure}

\subsection{Representation}
We treat the principal directions and stresses as a double orthogonal line field $\Psi(p) = (\vec{u}(p),\vec{v}(p))$ where $\vec{u}$ and $\vec{v}$ define the minimum and maximum stress at each point of the surface, respectively. 
Note that only the directions and sizes of $\vec{u},\vec{v}$ are relevant to $\Psi$, not their orientations. 
Since $\vec{u}$ and $\vec{v}$ are orthogonal, we decouple the scalar and directional information and represent $\Psi$ as a triple $(\vec{u}_n,d,a)$, where $\vec{u}_n$ is a unit-length vector parallel to $\vec{u}$, $d = |\vec{u}|$ is the maximum stress intensity (henceforth called \emph{density}), and  $a = |\vec{u}|/|\vec{v}|$ is the \emph{anisotropy} (see Figure \ref{fig:static_smoothing}).
This representation allows us to  better control the influence of $\Psi$ over the mesh generation process.

\subsection{Smoothing}
In most cases, the result of static analysis is not directly usable: the signal computed is often irregular with spikes of high stress and abrupt changes of direction in the line field, which are hard to handle during mesh generation. 
We smooth the line field $\vec{u}_n$ following the approach of Bommes et al.\ \shortcite{Bommes:2009}, modified as in \cite{Panozzo:2012}, see  Formula 5.
In short, we trade-off smoothness and faithfulness to the original line field, weighting the second term with anisotropy: we preserve those portions of field where there is a significant difference between the magnitudes of the two stress vectors, while obtaining a smoother field elsewhere.   

We also enforce that the two scalar signals $a$ and $d$  satisfy Lipschitz condition, i.e., $|d(p)-d(p+\vec{\varepsilon})| < L |\vec{\varepsilon}|$, with $L$ approximately equal to the diameter of the smallest Voronoi region we expect to obtain. 
%This is necessary to avoid a too fast variation of the signal.
This corresponds to a form of smoothing of the two scalar signals, which is performed through an upper saturation process that preserves the maxima of the function. 
%\TODO{Non sono riuscito a trovare una citazione sensata per questo}.
The results of smoothing are depicted on the lower side of Figure \ref{fig:static_smoothing}.

\subsection{Symmetrization}
\begin{figure}[t]
  \centering
  \includegraphics[width=8.5cm]{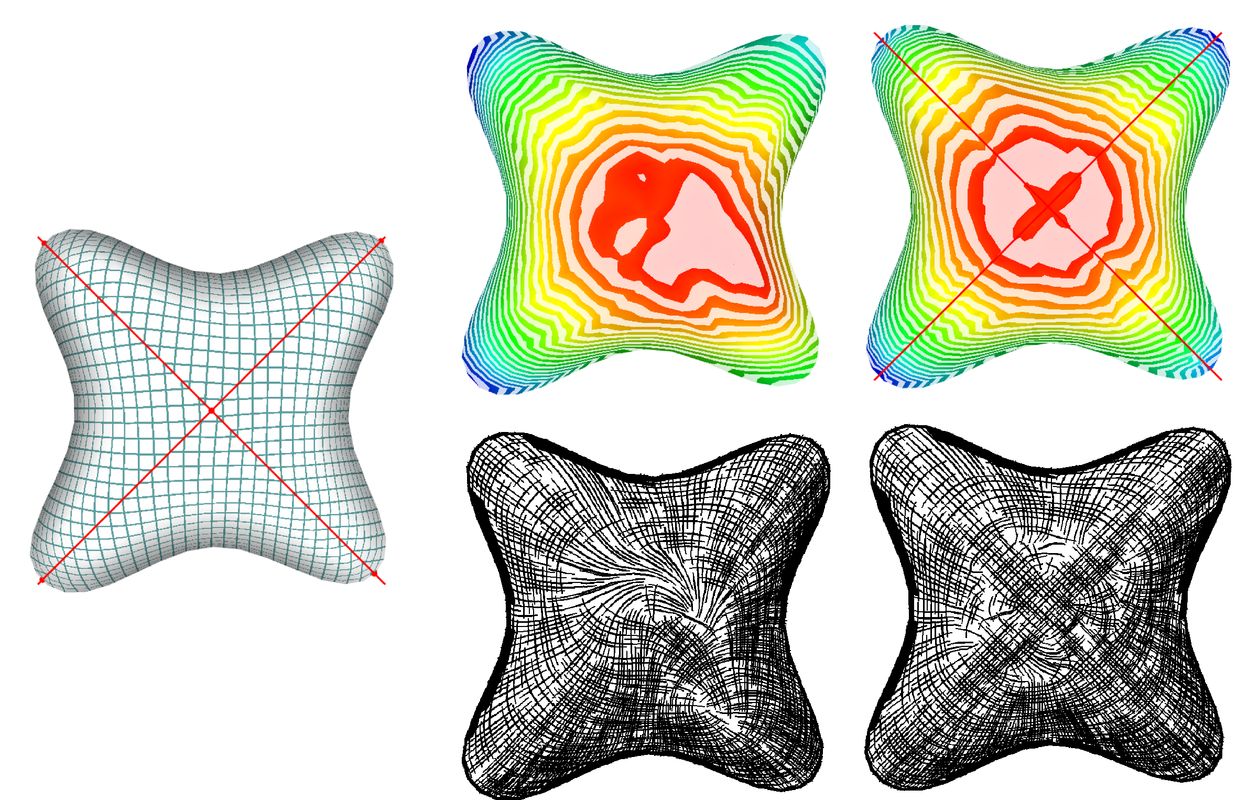}
  \caption{Symmetrization of $\Psi$ for the Lilium dataset: on the left the cross parametrization defined by two symmetry planes; on the center/right the density field (top) and the line field (bottom), before/after symmetrization.}
\label{fig:static_symmetry}
\end{figure}

Many architectural models present a few, sometime approximate, symmetry planes that should be preserved in the generated grid-shell. 
Assume we have one or more symmetry planes (shown in red in Fig.\ref{fig:static_symmetry}) that partition the mesh into regions.
We cross parameterize each symmetry region so that $\mu_{i,j}(p)$ be a cross-parametrization that maps a point $p$ of region $i$ onto its symmetric mate in region $j$.  
Cross-parametrizations are computed between adjacent regions in pairs and propagated about the center of symmetry. 
For two adjacent regions $i$ and $j$, we first cross-map corresponding points on their boundaries, exploiting the common boundary along the symmetry plane, plus symmetric corners that appear along intersections with other planes of symmetry and/or sharp corners on the boundary of the object. 
Then we compute a harmonic map for each region onto the same parametric domain, in such a way that symmetric points are mapped to the same point in parameter space. 
Finally, we compute a symmetric field $\bar{\Psi}$ by averaging it component-wise at all the corresponding points in the various regions (see Figure \ref{fig:static_symmetry}).
 
% !TEX root = main.tex
\section{Statics-aware ACVT} \label{sec:voronoi}
%The purpose of this section is to explain how w
%We use the metric induced by static analysis to drive the creation of an Anisotropic Centroidal Voronoi Tessellation (ACVT) that defines our grid-shell.

Let $S$ be a finite set of points, called \emph{seeds}, sampled in a metric space $\cal M$.
The Voronoi Diagram $V(S)$ is the partition of $\cal M$ into regions $V(S) = \{v(s),\ s\in S \} $ such that $v(s)$ is the portion of space closer to $s$ than to any other seed, with respect to the given metric on $\cal M$. 
The ACVT is the particular case of VD where the barycenter of each region $v(s)$ is coincident with the seed $s$ itself.
We are interested to the specific case of an ACVT defined over a bounded surface $\cal M$ embedded in $\mathds{E}^3$, with the metric induced by the stress tensor $\Psi$ defined in the previous section. 

\subsection{From general metric to Euclidean metric}
\begin{figure}[t]
  \centering
  \includegraphics[width=0.9\linewidth]{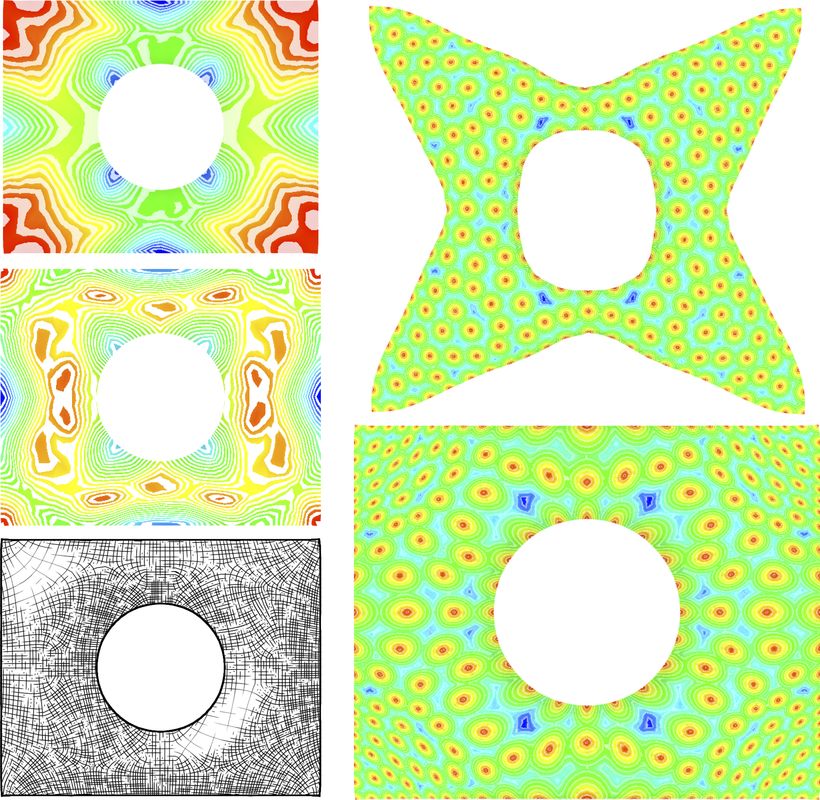}
  \caption{Density, anisotropy and directional field of the British Quad dataset (left); the resulting deformed domain mesh (top right) and the corresponding undeformed domain (bottom right) with the seeds of the CVT and their distance field.}
  \label{fig:deform}
\end{figure}

We interpret $\Psi$ as a frame field and we apply the method described in \cite{Panozzo:2014} to transform the metric induced by $\Psi$ into a Euclidean metric on a deformed surface $\cal M'$.
The metric induced by $\Psi$ on $\cal M$ is given by symmetric tensor $g_{\Psi}= \mathbf{W}^{-T}\mathbf{W}^{-1}$, with
\[\mathbf{W}=\left[ 
\begin{array}{cc}
d & 0\\
0 & \frac{d}{a}
\end{array}
\right] \] 
where $d$ and $a$ are the density and anisotropy described previously, and matrix $\mathbf{W}$ is expressed at each point in a local coordinate system aligned with $\Psi$. 
The metric becomes locally Euclidean if the underlying space in the neighborhood of each point $p$ is deformed by $\mathbf{W}^{-1}$ computed at $p$.
We evaluate $\mathbf{W}$ at each triangle of the input mesh $\cal M$, and we resolve an optimization problem that tends to deform each triangle $t$ to its ideal shape to make the metric Euclidean over $t$. 
See \cite{Panozzo:2014} for further details, and Figure \ref{fig:deform} for an example.

%In spite of the Lipschitz saturation, t
The density and anisotropy fields in input may span large intervals %contain drastic variations, 
which are not always desirable for designing a grid-shell.
We let the user adjust the desired variation of density and the desired amount of anisotropy over the surface, by introducing two parameters 
$D, A\geq 1$ and rescaling the $d$ and $a$ fields in the intervals $[1,D]$ and $[1,A]$, respectively, prior to computing deformation.
We show in Section \ref{sec:results} how such parameters can be used to fine tune the statics as well as the aesthetics of the grid-shell.
 
In order to improve the accuracy of subsequent computations, 
%Since we discretize CVT computation, by letting Voronoi seeds to be placed only at vertices of $\cal M'$, 
we refine the input mesh as follows, by subdividing edges that become too elongated under deformation.
We set a threshold $q$ for the maximum allowed length of an edge of $\cal M'$ (see next subsection about setting the value of $q$).
After deformation, we split all edges whose length exceeds $q$, together with their incident triangles, by midpoint subdivision. 
We estimate $\Psi$ at the centers of new triangles by interpolation, and we deform $\cal M'$ back to obtain a refined version of $\cal M$.
We iterate between deformation and refinement until all edge lengths are below $q$.

%Once we have the deformed domain $\cal M'$, we proceed to compute a CVT on it, then we transform the resulting tessellation back to $\cal M$.

\subsection{Seed Sampling}
We initialize the placement of seeds on $\cal M'$ by Poisson sampling \cite{Corsini:2012}, using a given radius $R$ of Poisson disks, which sets a user-defined sampling density, and placing seeds at vertices of $\cal M'$.
We adapt the refinement of $\cal M'$ to the desired sampling density by setting $q=R/5$ in the previous step. 
This value has been found experimentally to allow for a rather uniform distribution of seeds and good approximation of the VD.

Since we are dealing with a bordered domain and we want some seeds to remain on the border, we proceed as follows:
\begin{enumerate}
\item We first insert the vertices corresponding to sharp corners on the border of $\cal M$ into the set of seeds;
\item Then we sample just the border of $\cal M'$, constrained to the sharp corners;
\item Finally we sample the interior of $\cal M'$, constrained to the seeds inserted in the previous two steps.
\end{enumerate}

\subsection{Lloyd relaxation}
A CVT is computed through a standard iterative process known as Lloyd relaxation:
given a set of seeds, their VD is computed, then each seed is displaced to the centroid of its cell, and the process is repeated until convergence.

We compute a discrete approximated VD, which is sufficient to our purposes: the region of each seed $s$ is in fact the collection of vertices of $\cal M'$ that lie closer to $s$ then to the other seeds, according to an approximated geodesic distance computed with the fast method of Campen et al.\ \shortcite{Campen:2013}.

A crucial step of relaxation is the computation of centroids.
Given seed $s$ and its Voronoi region $v(s)$, at each iteration we choose the vertex inside $v(s)$ that minimizes the sum of the squared distances from all the other vertices in the region. 
Our approach is similar to the one in \cite{Valette:2004}, but it is based on a simpler, direct, linear computation: 
for each region we compute the quadric function $Q_s$ returning the sum of the squared distances from all the vertices in the region \cite{Garland:1998} and we evaluate $Q_s$ for the minimum over $v(s)$.

In order to preserve the boundary, seeds at sharp corners remain still during relaxation; while the other seeds on the boundary are displaced only along the boundary itself, moving a seed each time at the midpoint of its 1D Voronoi region; the boundary is relaxed at each iteration before relaxing the internal seeds.

The CVT is extracted easily from the discrete VD as follows: we set a Voronoi vertex $v_t$ at each triangle $t$ of $\cal M'$ whose vertices belong to three different Voronoi regions, by locating $v_t$ with barycentric coordinates on $t$ weighted through the distances of vertices of $t$ from their related seeds; and we connect pairs of Voronoi vertices that belong to the border of the same pair of regions.  
Finally, the ACVT of the original surface $\cal M$ is obtained by applying the reverse deformation to the vertices of the CVT of $\cal M'$, in order to bring them back to the surface of $\cal M$.

%======================= POLYGON REGULARIZATION =============================
\begin{figure}[t]
  \centering
  \includegraphics[width=2.5in]{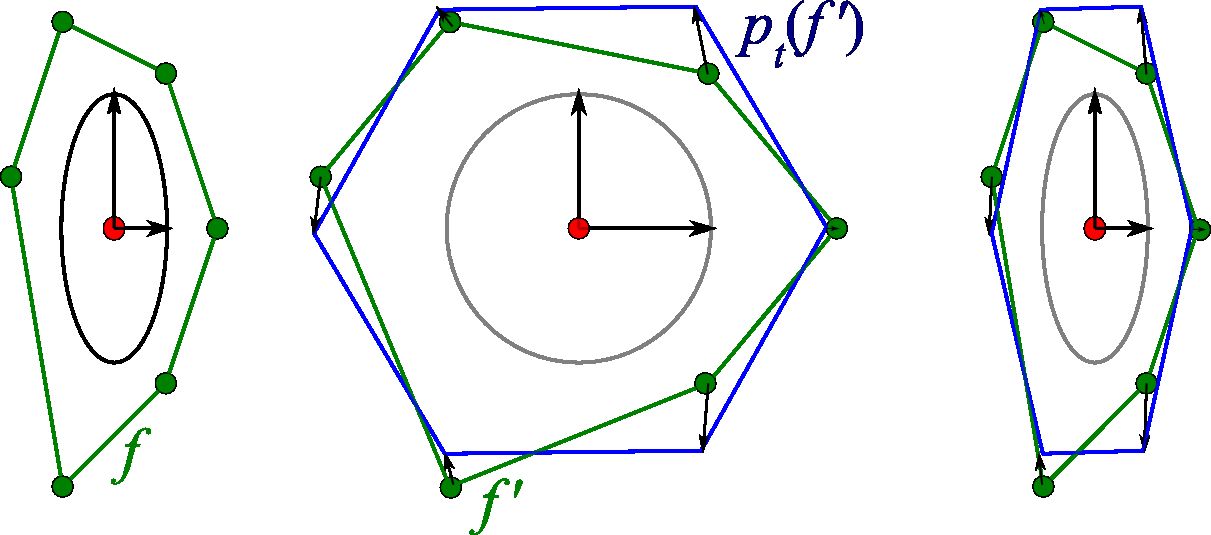}
  \caption{A single face $f$ of the ACVT with the eigenvectors resulting form PCA (left); the un-stretched polygon $f'$ with the aligned target polygon $p_t(f')$ (middle); and the computed displacement vectors in the original space (right).}
  \label{fig:regularization}
\end{figure}

%======================= SHAPE-UP COMPARISON =============================
\begin{figure}[t]
  \centering
  \includegraphics[width=1.45in]{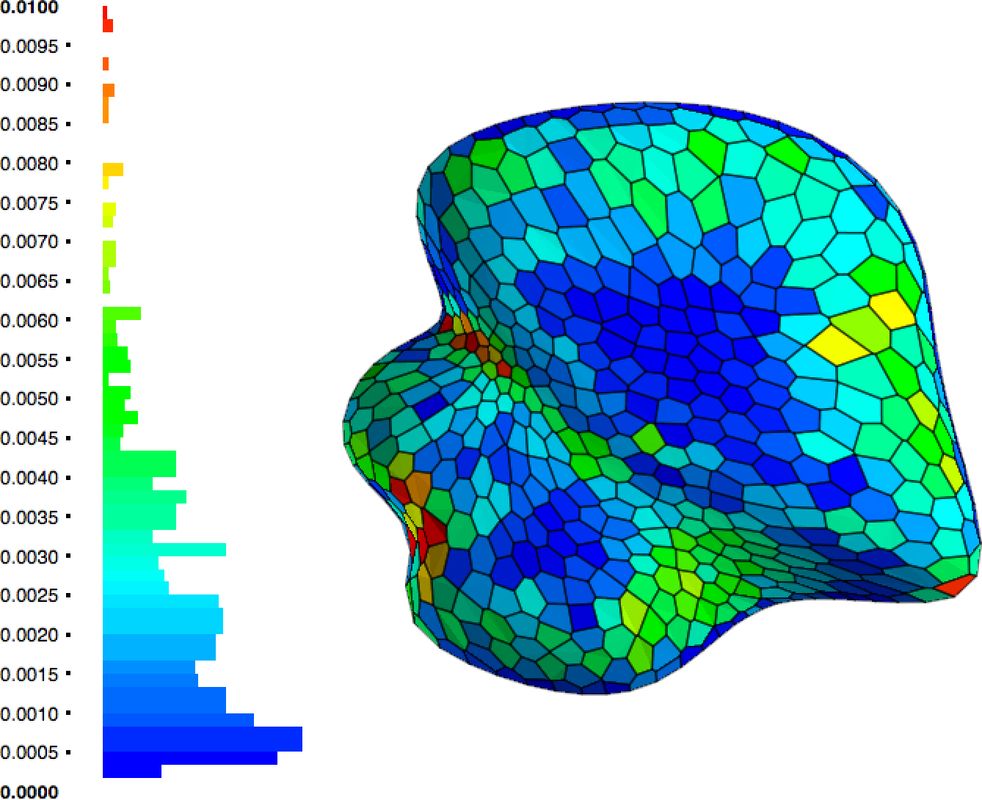}
  \includegraphics[width=1.45in]{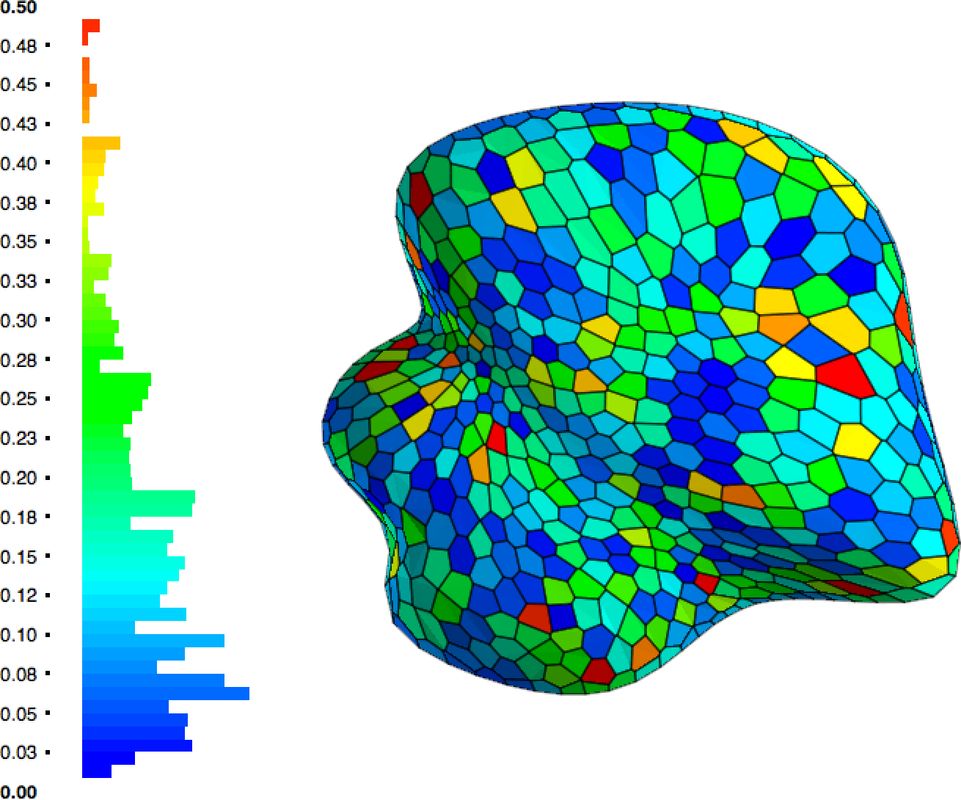}\\
 ~ \\
  \includegraphics[width=1.45in]{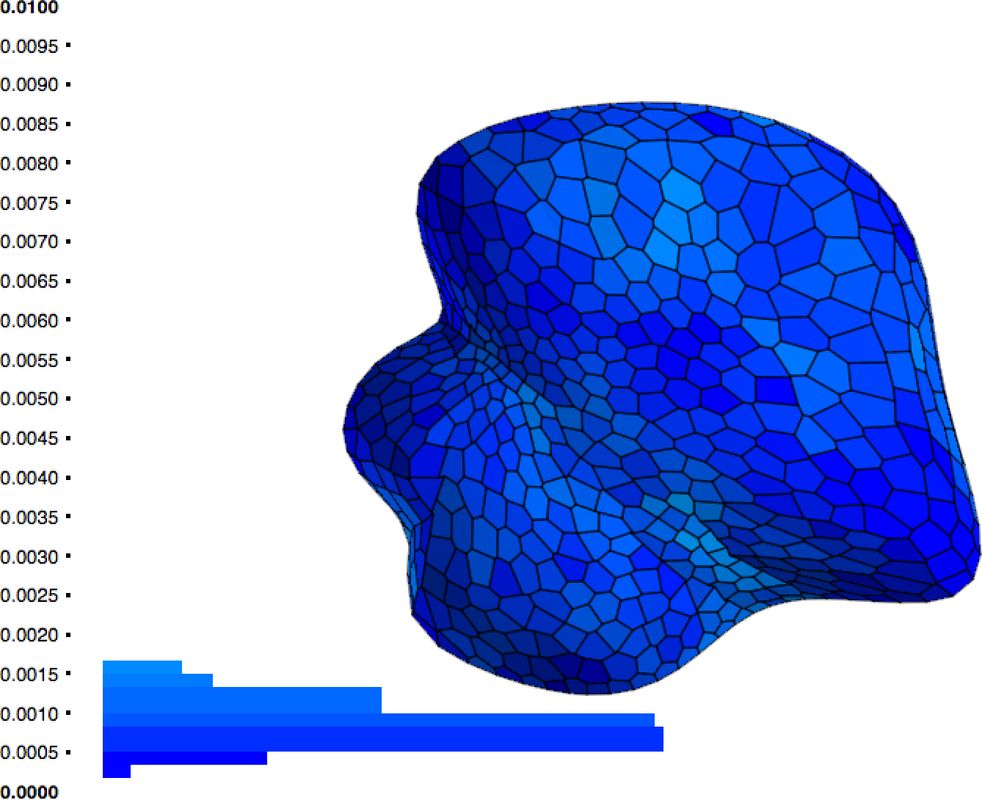}
  \includegraphics[width=1.45in]{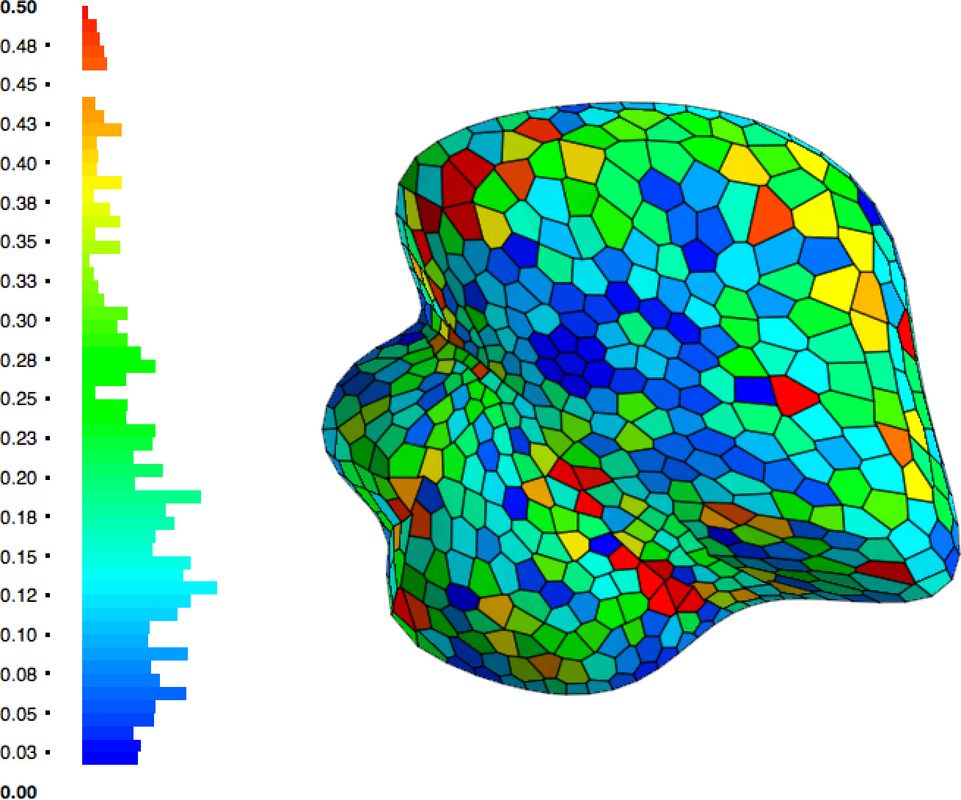}\\
 ~ \\
  \includegraphics[width=1.45in]{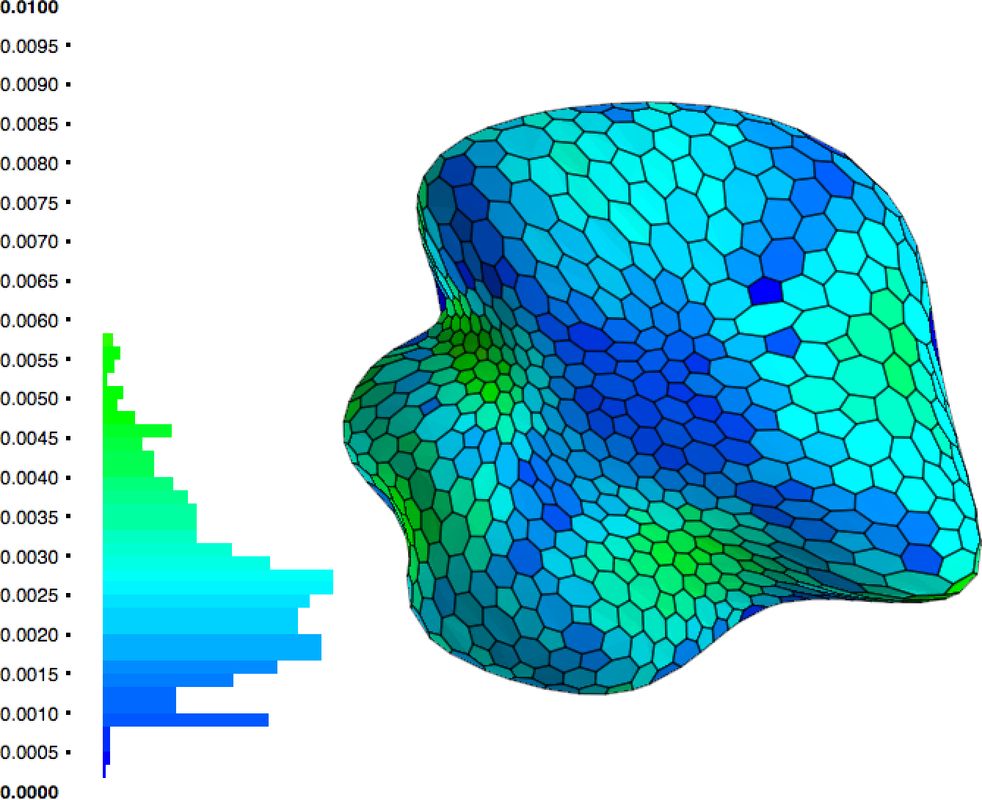}
  \includegraphics[width=1.45in]{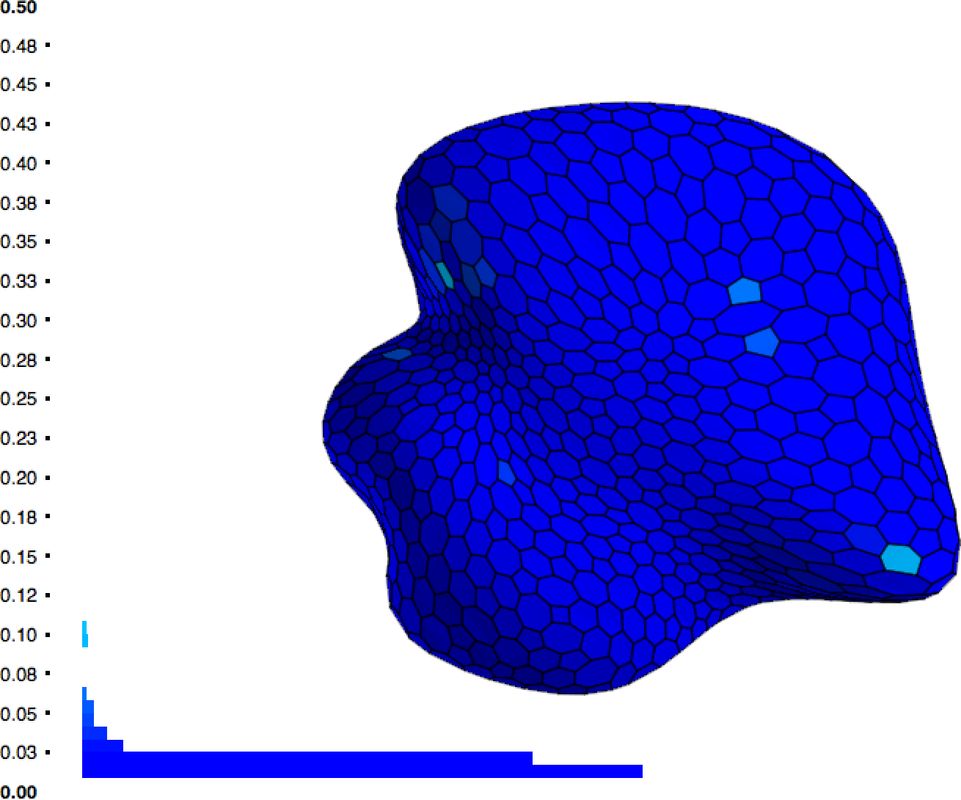}
  \caption{The effects of the regularization process on planarity (left) and regularity (right). The initial  ACVT (top), optimized for planarity with Shape-Up (middle) and regularized using our procedure (bottom).}
\label{fig:regular_comp}
\end{figure}

\subsection{Regularization}
%While CVT's ares usually known for their regular structure, we have found that in many cases more regularity in the final polygonal regions is needed. 
In order to improve the aesthetics, as well as the planarity of faces of the ACVT, 
%The main idea is that we want to improve the regularity of the generated polygonal shapes, that is, we want that each $v(s)$ is 
we optimize their shape to make them as similar as possible to stretched regular polygons. 
To this aim, we adopt a framework similar to \cite{Bouaziz:2012}, where we alternate {\em per-polygon} and {\em per-vertex} fitting steps.

In the per-polygon step, for each face $f$, we first perform a Principal Component Analysis to evaluate how much $f$ is stretched with respect to a regular polygon.
Then we compute a new polygonal region $f'$ corresponding to $f$ deformed (i.e., un-stretched) according to the two lowest rank eigenvectors of the PCA.
Next, we define a {\em target} regular polygon $p_t(f')$ having the same number of edges and equal perimeter as $f'$; 
then, using \cite{Besl:1992}, we rigidly align  $p_t(f')$ with $f'$; finally, we stretch the oriented polygon $p_t(f')$ back through the reverse deformation that was applied to $f$, and we use the vertices of  this stretched regular polygon as target positions to displace the vertices of $f$. 
Figure \ref{fig:regularization} shows the steps of this process for a single face. 

In the per-vertex fitting step, for each vertex $v$ independently, we evaluate the position minimizing the sum of squared distances from all the target positions specified for $v$ by its incident faces. 
We use a damping factor for improving convergence of this procedure. 

An interesting side effect of this regularization procedure is that it tends to make the length of edges more uniform, so that %at the end of the process 
the areas of faces will vary according to the number of sides of polygons. 
From an aesthetic point of view, this situation matches the look of the Archimedean class of semi-regular polyhedra. 

Given the similarity of this optimization approach with \cite{Bouaziz:2012}, we have also compared our results with the planarization approach presented in that paper. 
Figure \ref{fig:regular_comp} shows our approach in comparison with the initial ACVT and with the result of Shape-Up planarization.
As expected, Shape-Up achieves better planar faces, while our algorithm achieves a much better regularity of faces, hence better aesthetics.
Planarity is usually measured as distance between diagonals divided by average edge \cite{Tang:2014}. Unfortunately, this measure is not directly applicable in our case, since diagonals are ambiguous for polygons with an odd number of edges. 
Then, we  generalize the measure of planarity as the average distance of vertices to the best fitting plane divided by half perimeter. 
Regularity of a face is measured as the sum of squared distance of its vertices to their target positions, divided by its area.

%As shown in Section \ref{sec:results}, r
Regularization also slightly improves the overall structural properties of the grid-shell structure (10\% on average in our experiments).
The more uniform length of edges resulting from regularization is also an advantage during production. 

\subsection{Symmetry}
\begin{figure}[htcb]
\centering
\begin{tabular}{c c} 
\includegraphics[width=0.43\columnwidth]{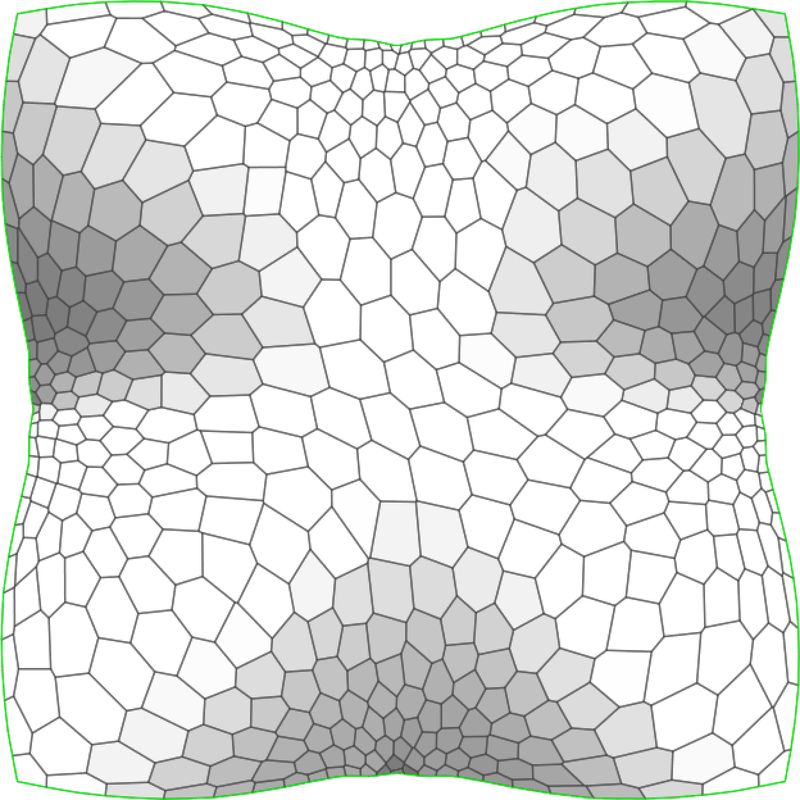}&
\includegraphics[width=0.43\columnwidth]{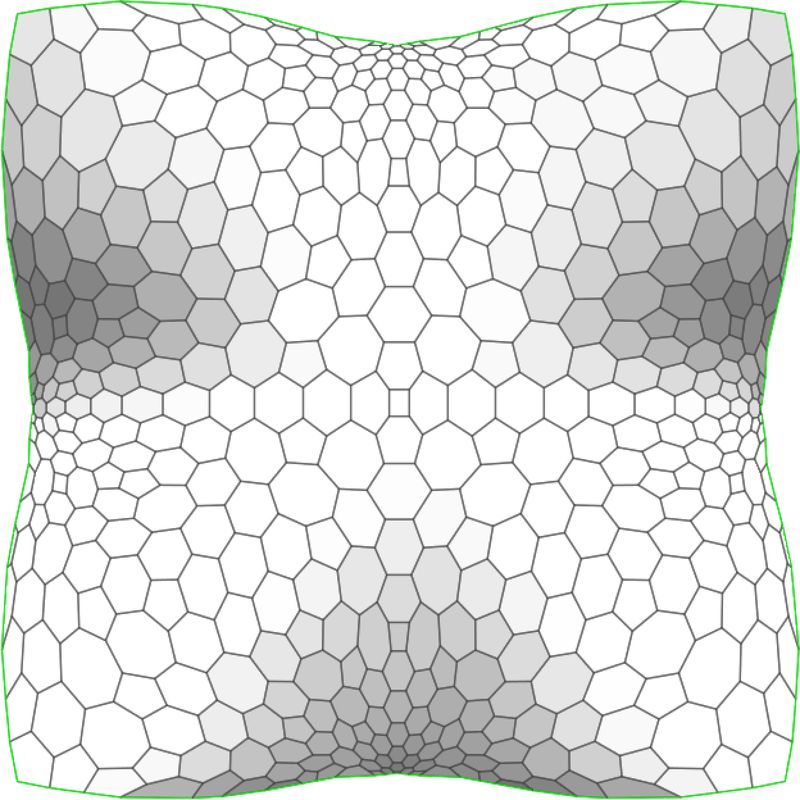}\\
Non Optimized & Optimized\\
$\lambda=2.92$  $\delta=33.48$ & $\lambda=3.05$  $\delta=43.10$
\end{tabular} 
	\caption{Comparison of non-optimized versus symmetrized and optimized tessellation. }
	\label{fig:ShellOptNoOpt}
\end{figure}

In order to improve the aesthetics of symmetrical structures, we use the same symmetry planes considered in Section \ref{sec:static} and we compute the ACVT just in one of the symmetric sectors.
Then we reflect the resulting tessellation to the other sectors, welding them at the regions of seeds placed along symmetry lines.
See Figure \ref{fig:ShellOptNoOpt} for a comparison between original ACVT and the optimized and symmetrized tessellation on the Shell dataset.

%===================== 	OPT VS NON-OPT =============================================

% !TEX root = main.tex

\section{Results} \label{sec:results}

Our method has been implemented in C++; static analysis has been performed by using the  \emph{GSA} Finite Element Analysis software \cite{GSA}, both on the input surface to obtain the stress tensor, and on the various grid-shells to test their behavior.
We have tested our method on several surfaces.
A summary of the datasets and related results are presented in Table \ref{tab:datasets}.

Overall, the running times for computing a grid-shell are negligible with respect to the times required to analyze results.
For instance, the models we have analyzed always took less than ten seconds to generate the stress tensor (with \emph{GSA}); and between one and ten minutes to build the grid-shell, depending on the number of iterations in the refinement-and-deformation step, which is the bottleneck of the pipeline. 
While the non-linear analysis of the result (again with \emph{GSA}) took over one day for the largest model analyzed.

%In the following, w
We present experiments that show the characteristics of our grid-shells in terms of statics, as well as some comparisons with grid-shells that are obtained with state-of-the-art methods in architectural geometry, or correspond to real-world architectures. 
All structures are assumed to be made of steel, consisting of solid bars with a diameter of 37 millimeters; and the load is distributed uniformly on the whole surface, i.e. each node gets a load that is proportional to the area of  its incident faces.
We evaluate the following measures, which are most relevant in structural engineering \cite{Meek}:  
\begin{itemize}
\item The \emph{non-linear buckling multiplier} $\lambda$, 
which measures the ability of a structure to support a load equal to a multiple of its weight before collapsing;
this attribute measures the robustness of the structure and it should be ideally maximized;
\item The \emph{nodal displacement} $\delta$,
which measures the maximum distance with respect to the reference shape when the structure is standing under serviceability load;
this attribute measures the degree of deformability of the structure and it should be ideally minimized.
\end{itemize}

%The first experiments aims to see how such physical characteristic variate with respect density and anisotropy interval. Then we 
\subsection{Tuning parameters}

Our method works on three parameters that must be set by the user: the threshold for density $D$, the threshold for anisotropy $A$, and the radius of sampling disks $R$. 
Comparative tests of static analysis require that different structures have the same weight and total length of beams %i.e., edges of the final mesh, 
%as it is common practice in structural engineering 
\cite{Malek:2013}.
We tolerate a 5\% of total length variation. 
%We relied only on optimized version of the tessellation since as in our experiments the optimization slightly improves the static performance of the grid-shell structure (see Figure \ref{fig:ShellOptNoOpt}).\\
In order to keep total length fixed, the number of faces must be decremented as density or anisotropy are incremented, and this is indeed possible by tuning parameter $R$.
In the following experiments, we thus take $D$ and $A$ as free parameters and we set $R$ as a constrained variable.
Then we test how the variation of density and anisotropy influence the buckling multiplier and nodal displacement. 

We have analyzed two funicular surfaces (datasets Shell and Paraboloid) and one light grid-shell surface (the dome designed by RFR Paris for the the Abbey of Neum\"{u}nster in Belgium). 
We vary density and anisotropy within a range that goes from 1 to 4, with unit step, for a total of 16 test cases per model. 
Due to the random sampling of seeds, the final meshes may result slightly different even for the same parameters. 
To disambiguate this randomness, we have performed three experiments for each parameter setting and we have averaged the results of static analysis (so the total number of experiments is in fact 48 per model).
 
\begin{figure}[h!]
\centering
\begin{tabular}{ @{}c@{} }

	\begin{tabular}{ @{}c@{}c@{} }

		\begin{tabular}{ @{}c@{}}
 			D=1\\ \begin{sideways} ... \end{sideways}
 		\end{tabular} &
	
		\begin{tabular}{ @{}cc@{}} 
 			A=1... &  ... A=4\\
			\includegraphics[height=0.215\columnwidth]{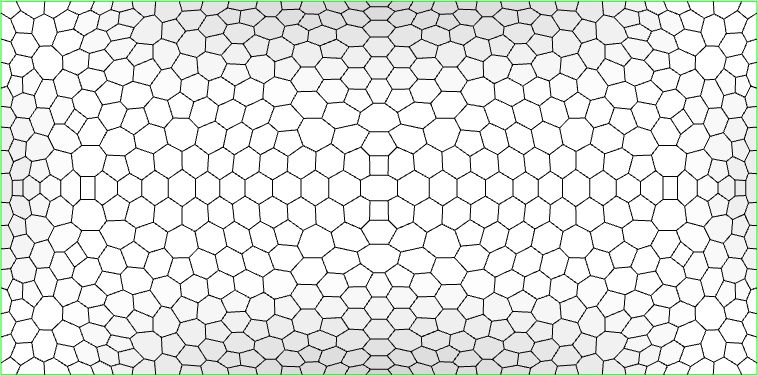}&
			\includegraphics[height=0.215\columnwidth]{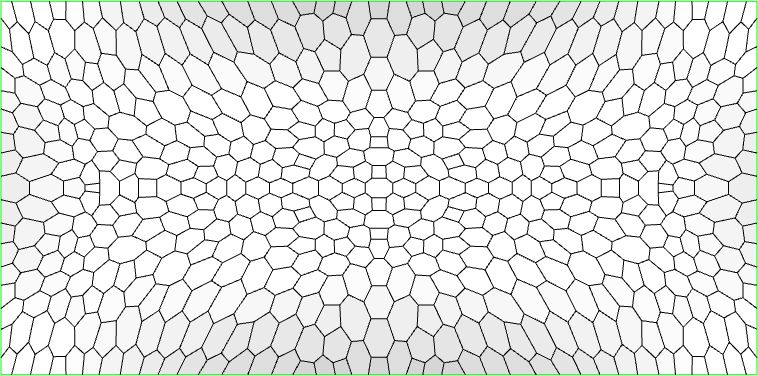}\\
		\end{tabular} \\
		
		\begin{tabular}{ @{}c@{}}
 			\begin{sideways} ... \end{sideways} \\ D=4
 		\end{tabular} &
		\begin{tabular}{ @{}cc@{}} 
 			\includegraphics[height=0.215\columnwidth]{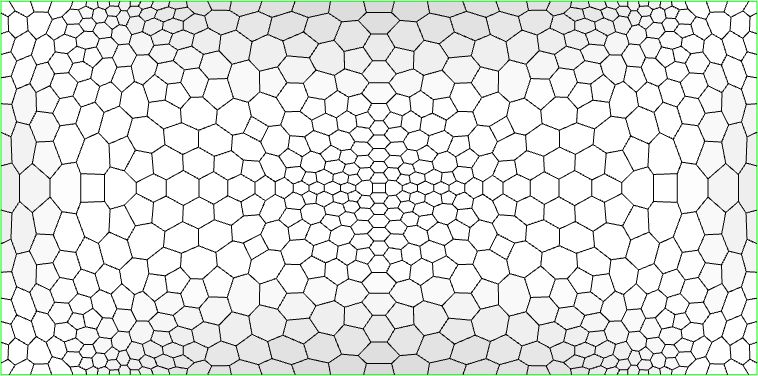}&
			\includegraphics[height=0.215\columnwidth]{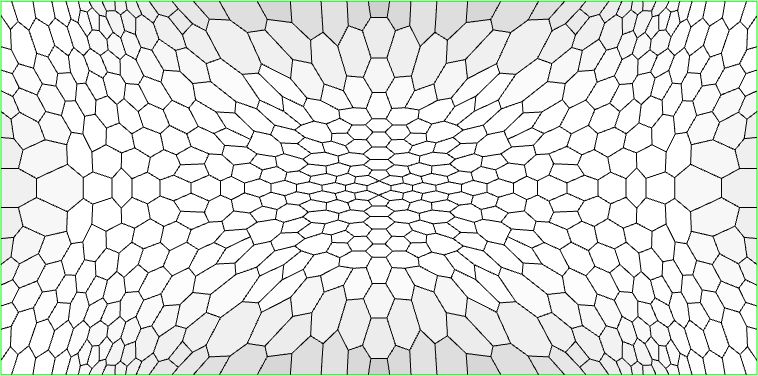}\\
		\end{tabular}  \\
	\end{tabular} \\
	
	\begin{tabular}{ @{}c@{}} 
 		\includegraphics[height=0.8\columnwidth]{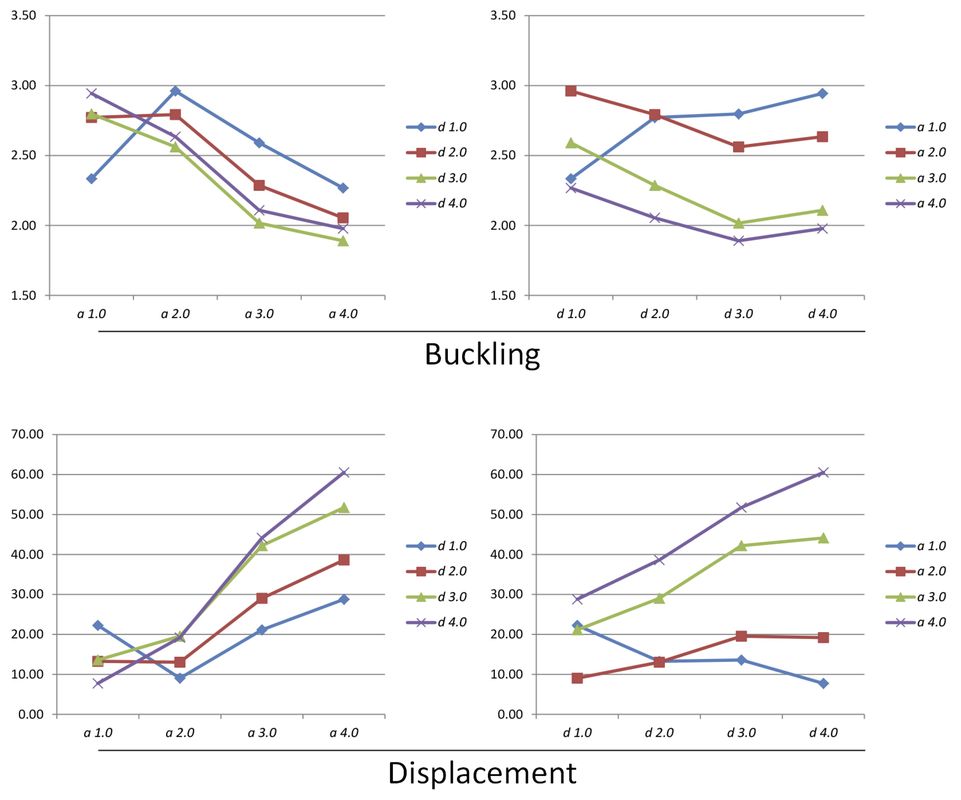}\\
	\end{tabular} 

\end{tabular} 
	\caption{4x4 test on the Neum{\"u}nster model. }
	\label{fig:NeumoisterAnalisis}
\end{figure}

\begin{figure}[h!]
\centering
\begin{tabular}{ @{}c@{} }

	\begin{tabular}{ @{}c@{}c@{} }

		\begin{tabular}{ @{}c@{}}
 			D=1\\ \begin{sideways} ... \end{sideways}
 		\end{tabular} &
	
		\begin{tabular}{ @{}cc@{}} 
 			A=1... &  ... A=4\\
			\includegraphics[height=0.41\columnwidth]{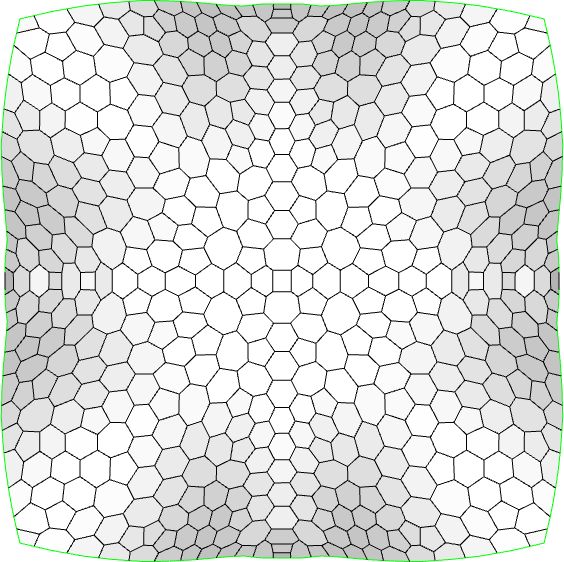}&
			\includegraphics[height=0.41\columnwidth]{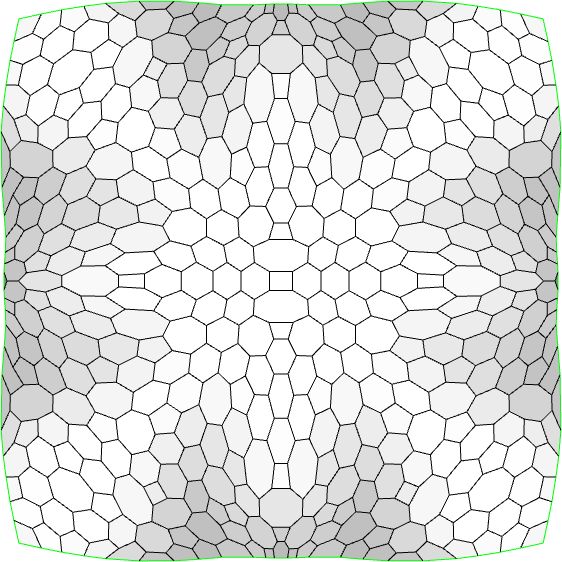}\\
		\end{tabular} \\
		
		\begin{tabular}{ @{}c@{}}
 			\begin{sideways} ... \end{sideways} \\ D=4
 		\end{tabular} &
		\begin{tabular}{ @{}cc@{}} 
 			\includegraphics[height=0.41\columnwidth]{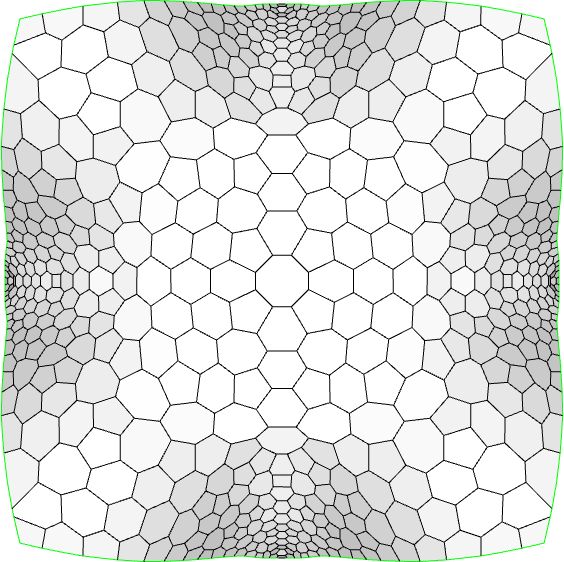}&
			\includegraphics[height=0.41\columnwidth]{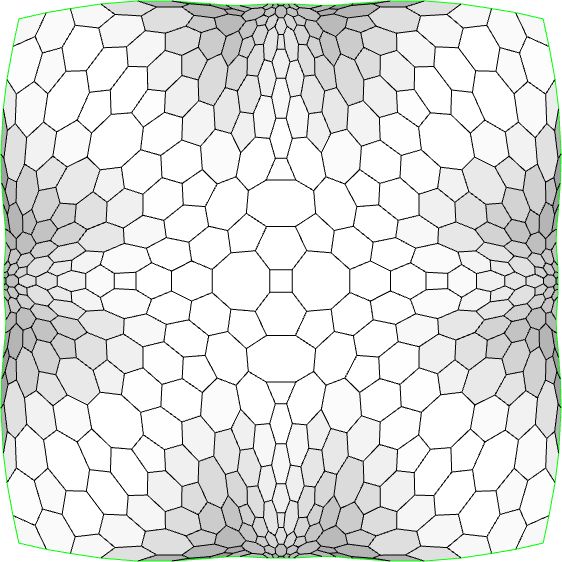}\\
		\end{tabular}  \\
	\end{tabular} \\
	
	\begin{tabular}{ @{}c@{}} 
 		\includegraphics[height=0.8\columnwidth]{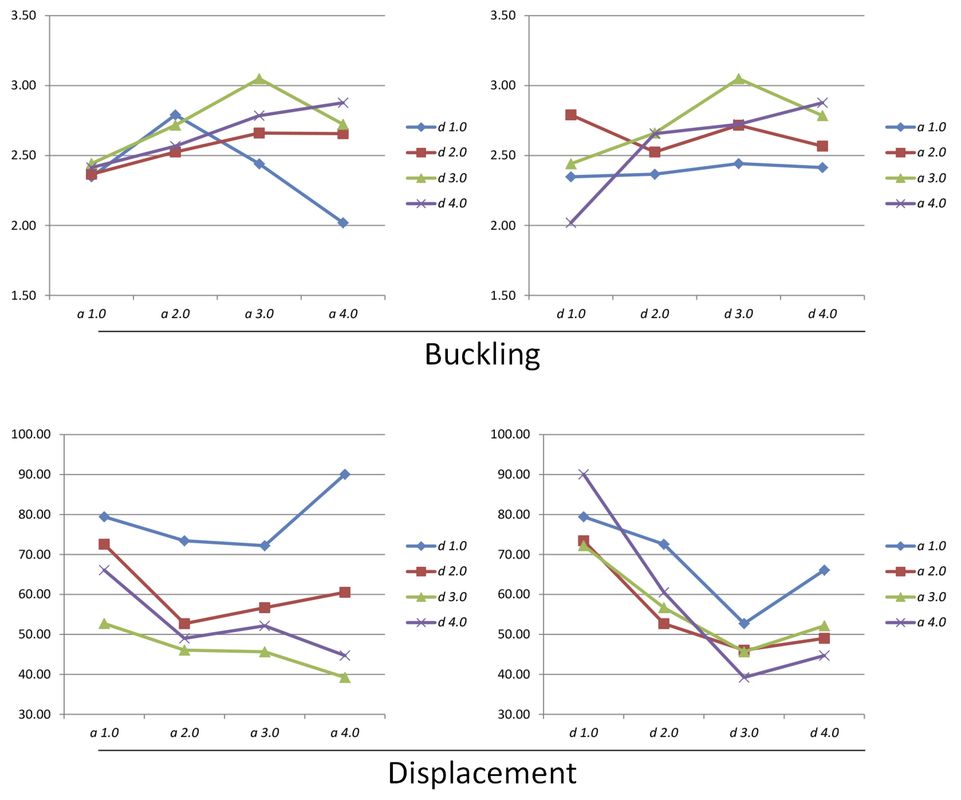}\\
	\end{tabular} 

\end{tabular} 
	\caption{4x4 test on the Shell model. }
	\label{fig:ShellAnalisis}
\end{figure}

\begin{figure}[h!]
\centering
\begin{tabular}{ @{}c@{} }

	\begin{tabular}{ @{}c@{}c@{} }

		\begin{tabular}{ @{}c@{}}
 			D=1\\ \begin{sideways} ... \end{sideways}
 		\end{tabular} &
	
		\begin{tabular}{ @{}cc@{}} 
 			A=1... &  ... A=4\\
			\includegraphics[height=0.42\columnwidth]{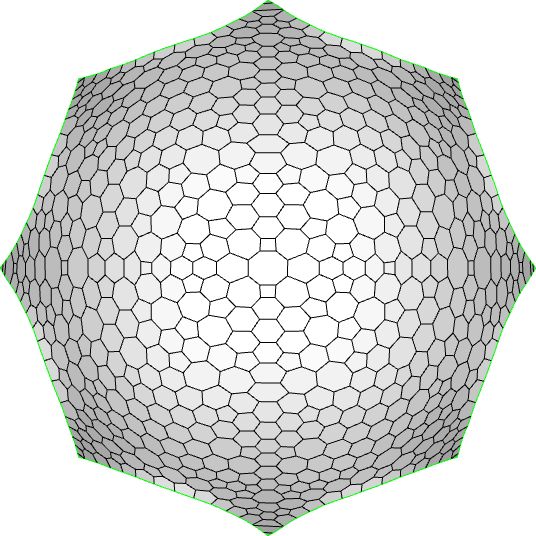}&
			\includegraphics[height=0.42\columnwidth]{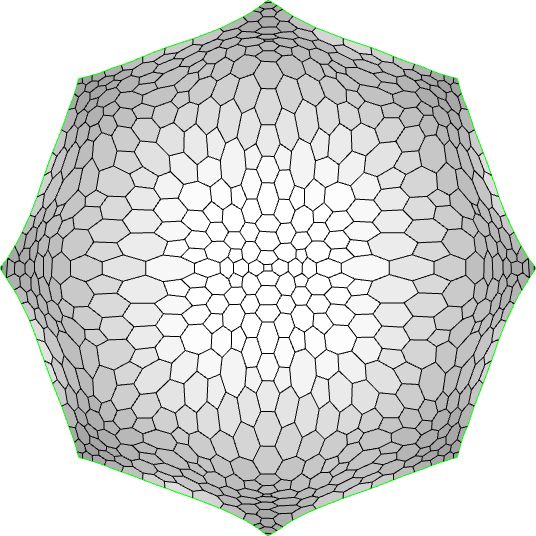}\\
		\end{tabular} \\
		
		\begin{tabular}{ @{}c@{}}
 			\begin{sideways} ... \end{sideways} \\ D=4
 		\end{tabular} &
		\begin{tabular}{ @{}cc@{}} 
 			\includegraphics[height=0.42\columnwidth]{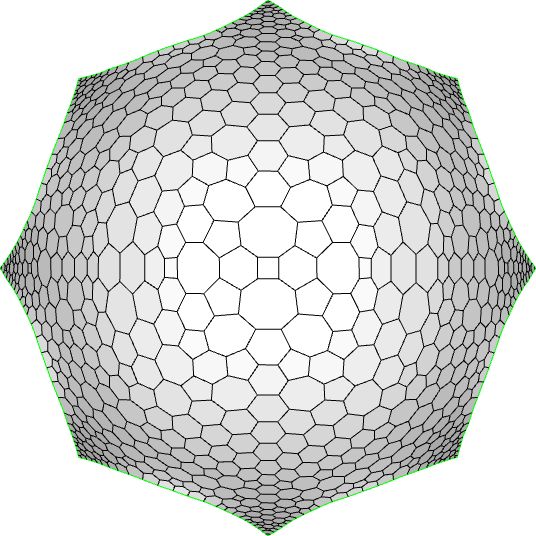}&
			\includegraphics[height=0.42\columnwidth]{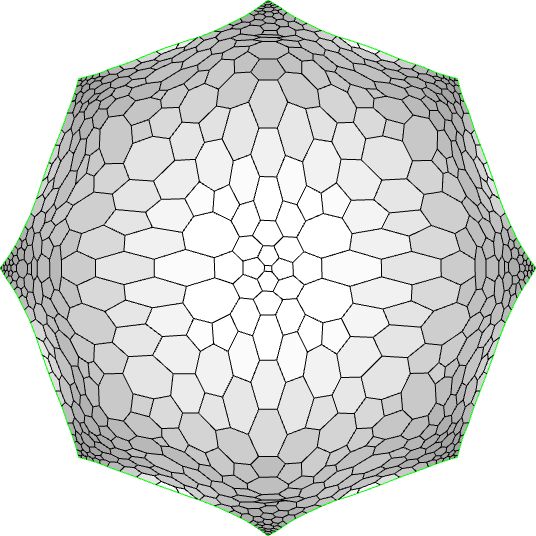}\\
		\end{tabular}  \\
	\end{tabular} \\
	
	\begin{tabular}{ @{}c@{}} 
 		\includegraphics[height=0.8\columnwidth]{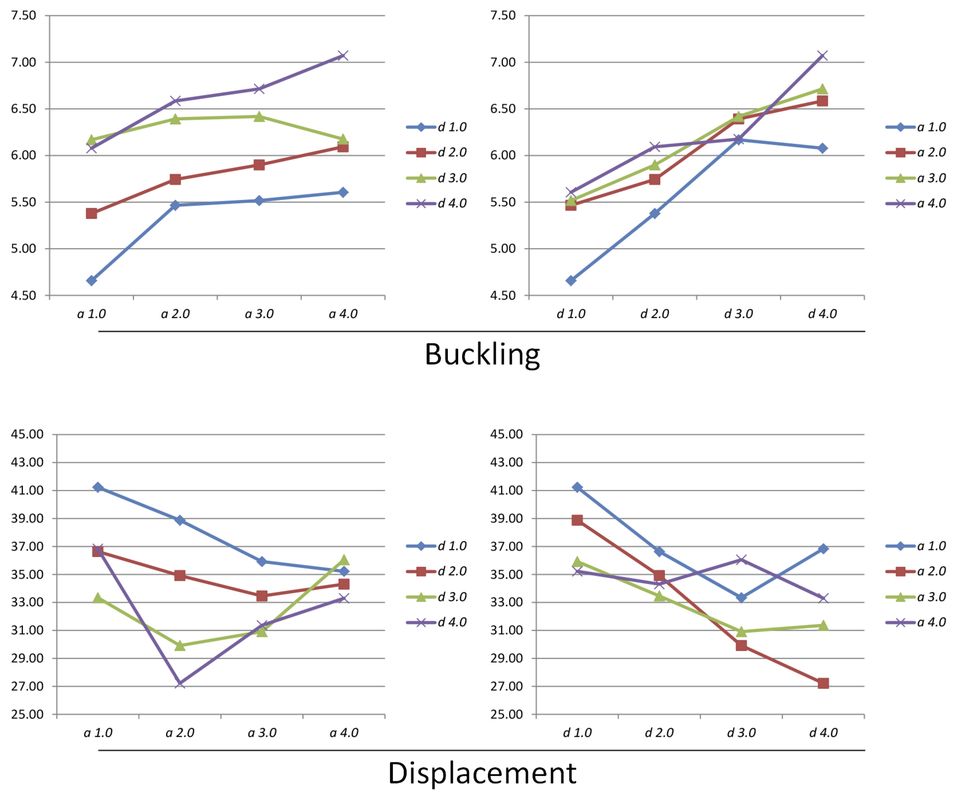}\\
	\end{tabular} 

\end{tabular} 
	\caption{4x4 test on the Paraboloid model. }
	\label{fig:ParaboloidAnalisis}
\end{figure}

\begin{figure*}[tcbp]
\centering
\begin{tabular}{ @{}c@{}c@{}c@{}} 
\includegraphics[width=0.66\columnwidth]{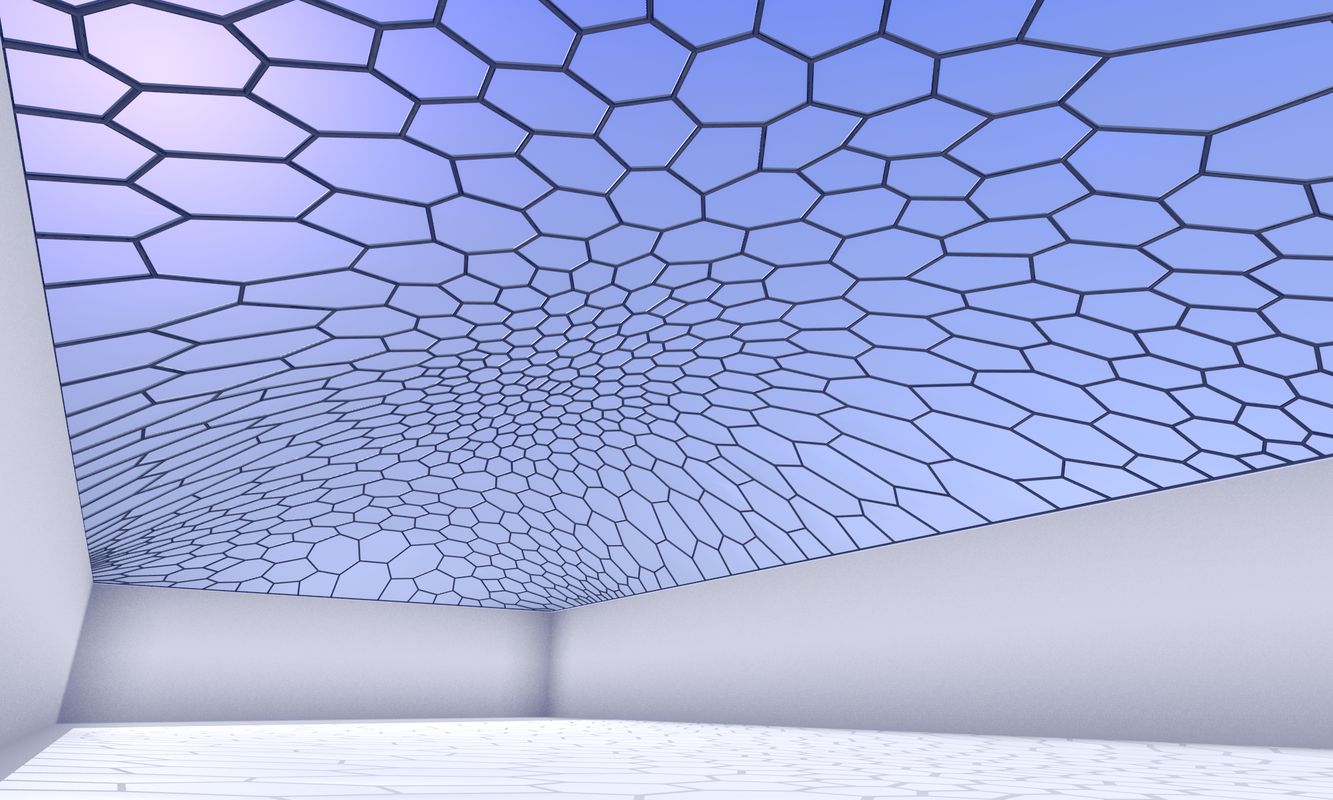}&
\includegraphics[width=0.66\columnwidth]{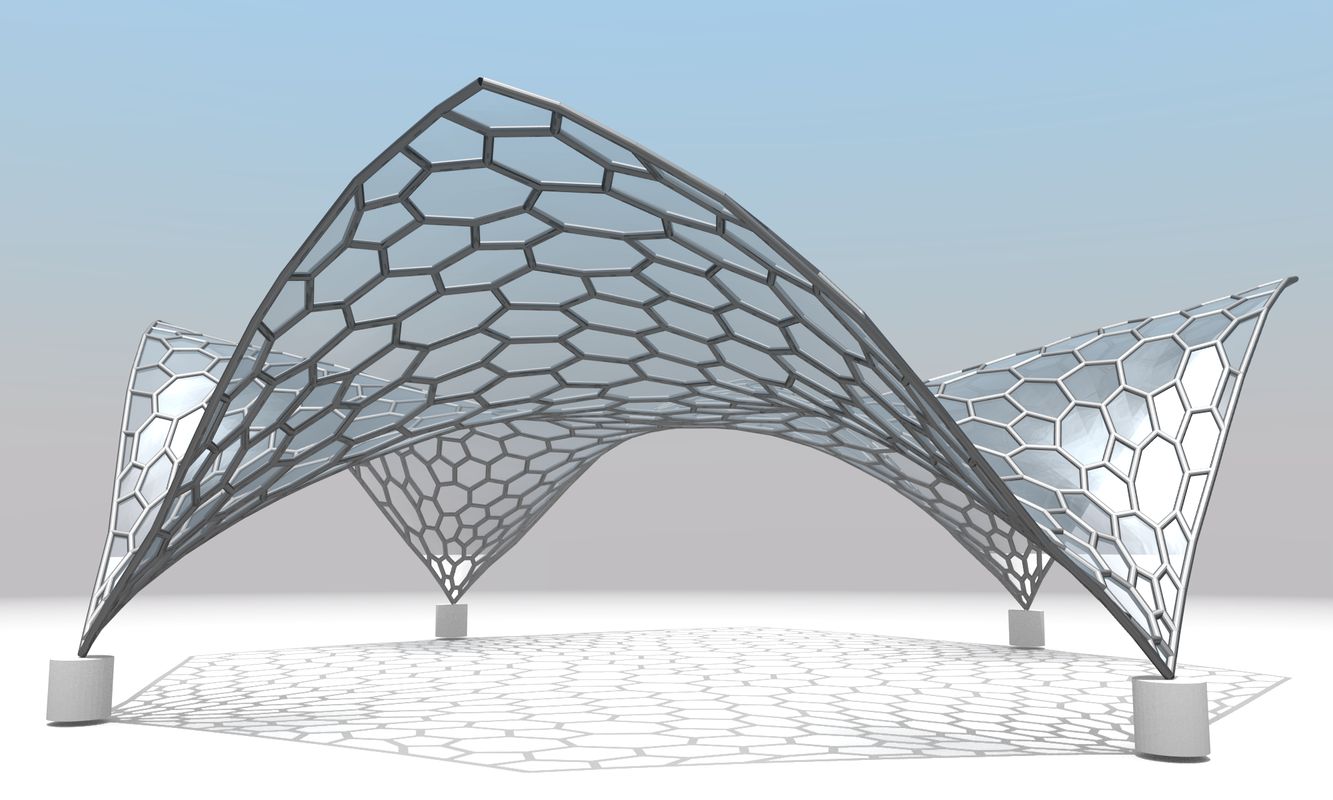}&
\includegraphics[width=0.66\columnwidth]{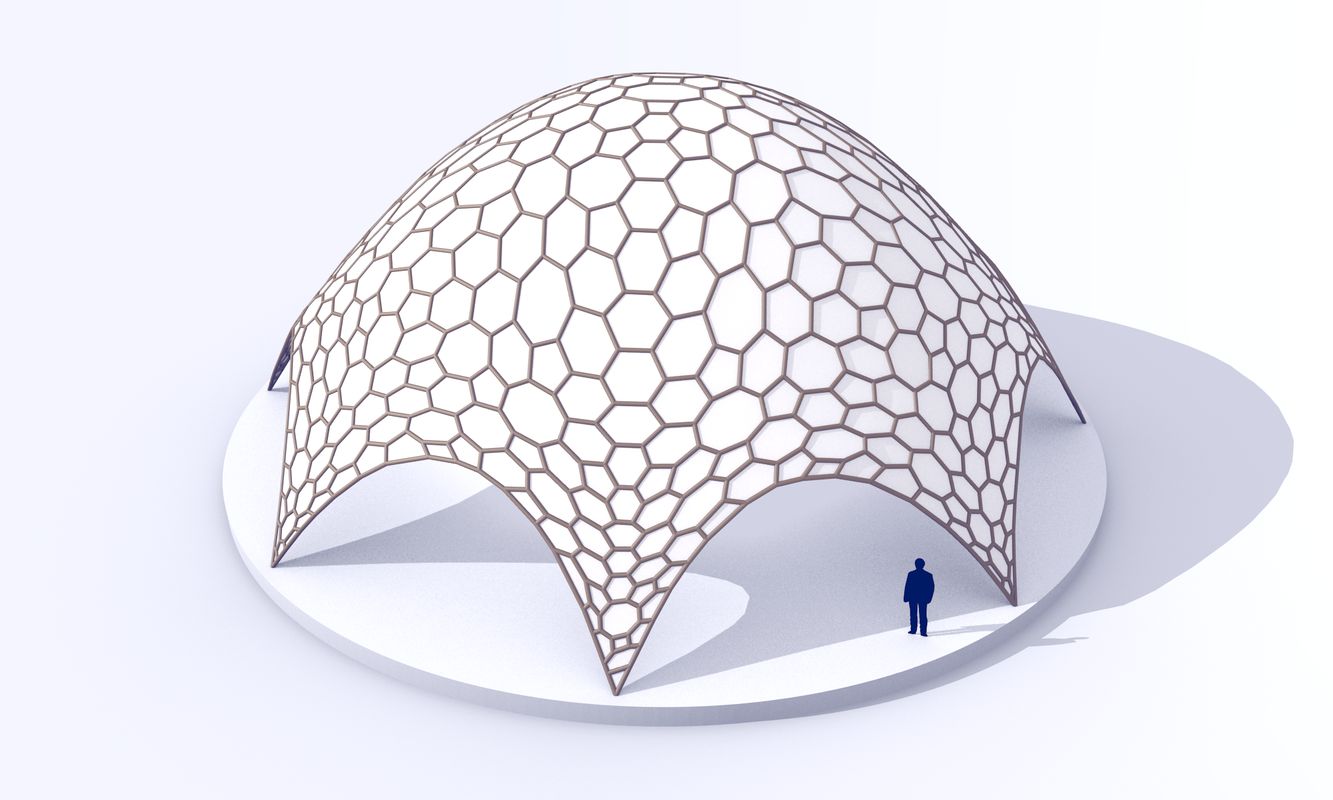}\\
\end{tabular} 
	\caption{The models used for parameter tuning tests. From the left: Neum{\"u}nster, Shell, Paraboloid.} %The last model shows the original model for the Neum\"{u}nster  gridshell. }
	\label{fig:RenderingTuning}
\end{figure*}

Some pictures illustrating the experiment are shown in Figures \ref{fig:NeumoisterAnalisis}, \ref{fig:ShellAnalisis} and \ref{fig:ParaboloidAnalisis} (top views and graphs), while rendering of the best performing models for each dataset are presented in Figure \ref{fig:RenderingTuning}.

The Neum\"{u}nster dataset achieves the highest buckling for $(D,A)=(2,1)$ and the lowest displacement for $(D,A)=(4,1)$, and the latter setting gives the best compromise. 
This suggests that for this kind of dataset, which is supported on the whole perimeter, density plays a relevant role in improving statics, while anisotropy does not help.
The Shell and Paraboloid datasets, on the contrary, rest on a small number of points. 
The Shell achieves the highest buckling for $(D,A)=(3,3)$ and the lowest displacement for $(D,A)=(3,1)$, and the former setting gives the best compromise.
The Paraboloid achieves the highest buckling for $(D,A)=(4,4)$ and the lowest displacement $(D,A)=(4,1)$, but  $(D,A)=(4,2/3)$ give the best compromise.
This suggests that for this class of surfaces variations in both density and anisotropy help improving statics.

%Even if we kept fixed the total amount of steel, It is, in general, a  tricky task to relate our density/anisotropy parameters to non-linear physical behaviour of a structure that is about to collapse. However, from this experiment we may observe that the best parameters are strictly related to the shape we are modeling.  
%However, we may notice that shapes that have a low number of constraints  has its own specific behaviour, it is 

%===================== 	TABLE OF DATASETS =============================================
\begin{table*}[tbp]
%\small
\centering
%\setlength{\tabcolsep}{4pt}
%\ra{1.1}
\begin{tabular}{l l r  r  r r r r}
\hline
  Dataset &  Model & \# Vertices  & \# Faces    & \# Edges & Total length (m) & $\lambda$ & $\delta$\\ 
\hline
  Aquadom	        & \cite{Vouga:2012}	& 1078	&  1004	& 	2074		& 3906	& 1.51	& 144.90\\  
 				& Quad (3,3) 		& 1293	&  1052	& 	2352		& 3938	& 2.86	& 54.04\\ 
				\rowcolor{LightCyan}
 				& Voronoi (3,3) 		& 2382	&  1177	& 	3752		& 3898	& 3.44	& 48.01\\ 
\hline
  Botanic		        & \cite{Tang:2014}	& 1121	&  1076	& 	2196		& 1989	& 1.00	& 271.49\\  
 				& Quad (3,3) 		& 1194	&  1039	& 	2232		& 2006	& 1.80	& 59.69\\ 
				\rowcolor{LightCyan}
 				& Voronoi (3,3) 		& 2436	&  1202	& 	3654		& 2018	& 1.76	& 91.59\\ 
\hline
  British Quad	        & \cite{Tang:2014}	& 1648	&  1568	& 	3216		& 4286	& 2.44	& 19.57\\  
 				& Quad (2,3) 		& 1987	&  1585	& 	3583		& 4145	& 2.32	& 24.23\\ 
				\rowcolor{LightCyan}
 				& Voronoi (2,3) 		& 3812	&  1974	& 	5868		& 4314	& 5.78	& 7.65\\ 
\hline
  British Tri		& Tri (real)			& 1746	&  3312	& 	4878		& 10267	& 9.62	& 2.6\\  
  				\rowcolor{LightCyan}
 				& Voronoi (2,3) 		& 3024	&  5110	& 	15332	& 10799	& 6.82	& 15.09\\ 
\hline
  Lilium		        & \cite{Vouga:2012}	& 1648	&  636	& 	3216		& 4286	& 1.73	& 61.43\\  
 				& Quad (3,4) 		& 1987	&  660 	& 	3583		& 4145	& 1.86	& 52.67\\ 
				\rowcolor{LightCyan}
 				& Voronoi (3,4) 		& 3812	&  695	& 	5868		& 4314	& 6.54	& 16.51\\ 
\hline
%  Neum{\"u}nster	& tri (original)		& 220	& 380	& 	541		& 967\\  
\rowcolor{LightCyan}
Neum{\"u}nster		& Voronoi (2,2) 		& 1252	&  571	& 	1602		& 975	& 1.71	& 10.35\\ 
\rowcolor{LightCyan}
Paraboloid		& Voronoi (4,2) 		& 1424	&  745	& 	2064		& 2156	& 6.59	& 27.22\\ 
\rowcolor{LightCyan}
Shell				& Voronoi (3,3) 		& 988	&  477	& 	1441		& 415	& 3.05	& 46.65\\ 
\hline
\end{tabular}
\caption{Statistics on datasets and results: for each dataset we show statistics on models taken for comparison and models built by us.
Models from \protect\cite{Tang:2014,Vouga:2012} are quad meshes. Note that British Tri and British Quad refer to different surfaces. 
Quad and Voronoi refer to our models of anisotropic quad meshes and ACVT, respectively, computed with parameters $(D,A)$. 
For each model we report: the number of vertices, faces and edges; the total length of beams in the model; the buckling factor $\lambda$; and the nodal displacement $\delta$.}
\label{tab:datasets}
\end{table*}
%===================== 	END TABLE OF DATASETS ==========================================

%===================================

%===================================

%===================================

%===================================

%\begin{figure*}[htcb]
%\centering
%\begin{tabular}{ @{}cccc@{}} 
%\textbf{British Museum} & \textbf{Botanic Garden} & \textbf{Lilium} & \textbf{Aquadom}\\
%	\includegraphics[height=0.35\columnwidth]{./images/voro_vs_quad/British_tang.jpg}&
%	\includegraphics[height=0.35\columnwidth]{./images/voro_vs_quad/Botanic_tang.jpg}&
%	\includegraphics[height=0.35\columnwidth]{./images/voro_vs_quad/lilium_vouga.jpg}&
%	\includegraphics[height=0.35\columnwidth]{./images/voro_vs_quad/Aquadom_Vouga.jpg}\\
%	\cite{Tang:2014}&\cite{Tang:2014}&\cite{Vouga:2012}&\cite{Vouga:2012}\\
%	\hline
%	\includegraphics[height=0.35\columnwidth]{./images/voro_vs_quad/British_miq.jpg}&
%	\includegraphics[height=0.35\columnwidth]{./images/voro_vs_quad/Botanic_miq.jpg}&
%	\includegraphics[height=0.35\columnwidth]{./images/voro_vs_quad/Lilium_miq.jpg}&
%	\includegraphics[height=0.35\columnwidth]{./images/voro_vs_quad/Aquadom_miq.jpg}\\
%	\multicolumn{4}{c}{Anisotropic Mixed Integer Quadrangulation}\\
%	\hline
%	\includegraphics[height=0.35\columnwidth]{./images/voro_vs_quad/British_our.jpg}&
%	\includegraphics[height=0.35\columnwidth]{./images/voro_vs_quad/Botanic_our.jpg}&
%	\includegraphics[height=0.35\columnwidth]{./images/voro_vs_quad/Lilium_our.jpg}&
%	\includegraphics[height=0.35\columnwidth]{./images/voro_vs_quad/Aquadom_our.jpg}\\
%	\multicolumn{4}{c}{Voronoi Grid Shell}\\
%\end{tabular} 
%	\caption{Comparison with quad meshes. } %The last model shows the original model for the Neum\"{u}nster  gridshell. }
%	\label{fig:VsQuads}
%\end{figure*}

\begin{figure}[htcb]
\centering
\begin{tabular}{ @{}c@{}c@{}c@{}} 
	\includegraphics[height=0.25\columnwidth]{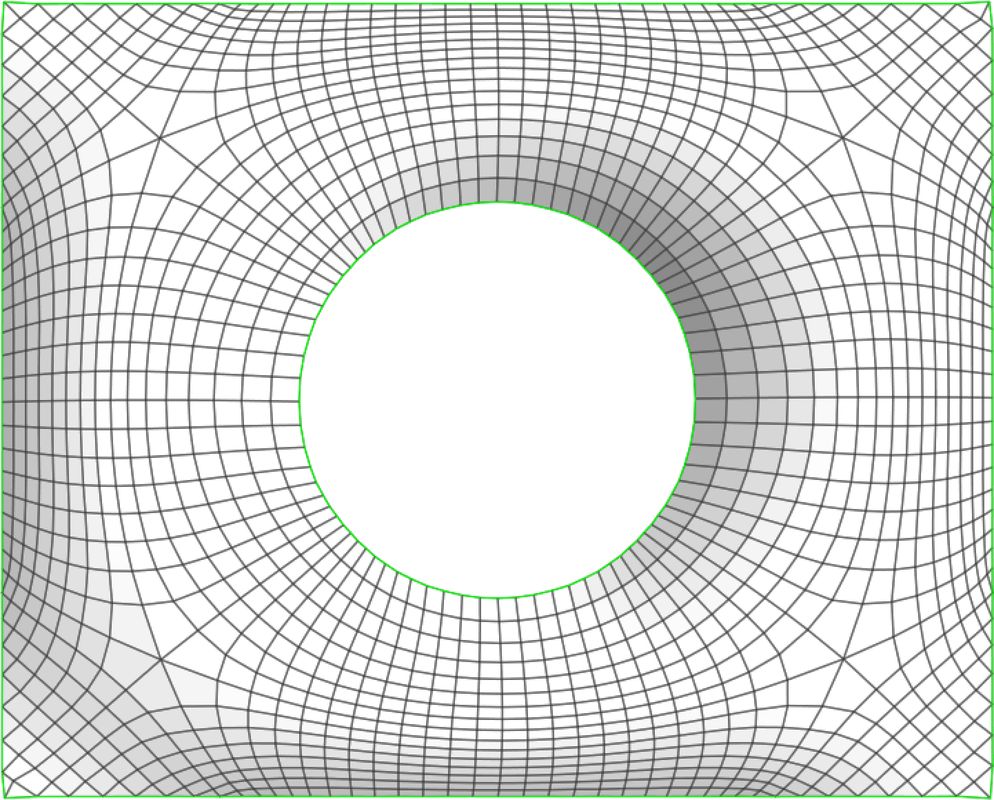}&
	\includegraphics[height=0.25\columnwidth]{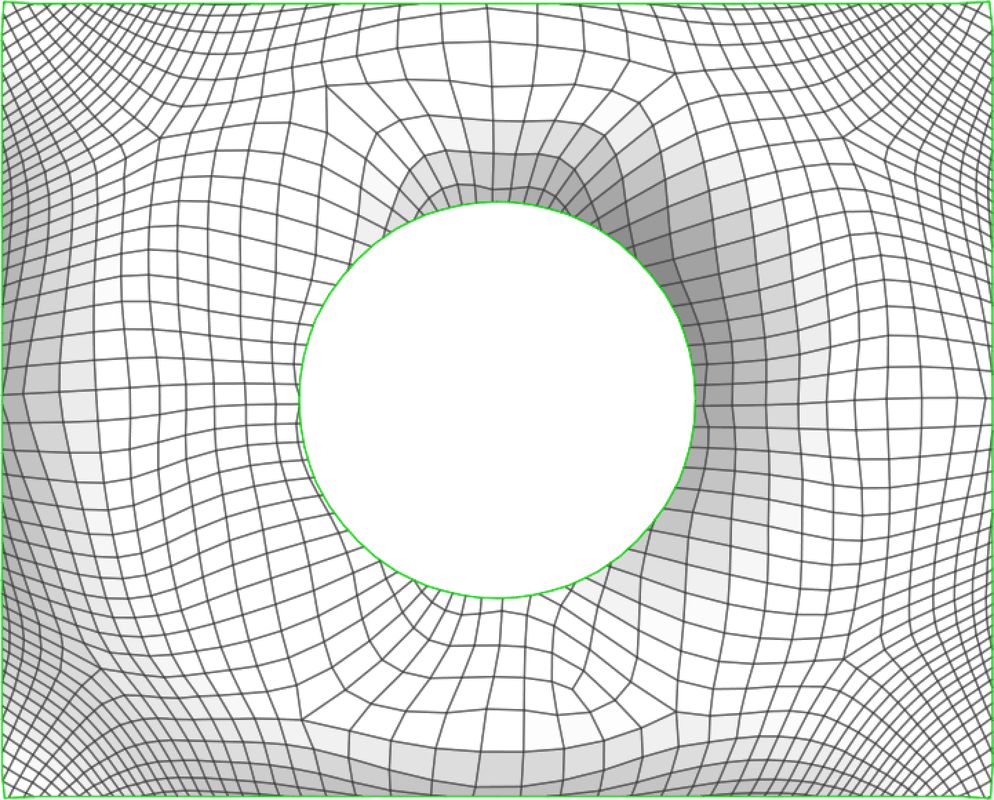}&
	\includegraphics[height=0.25\columnwidth]{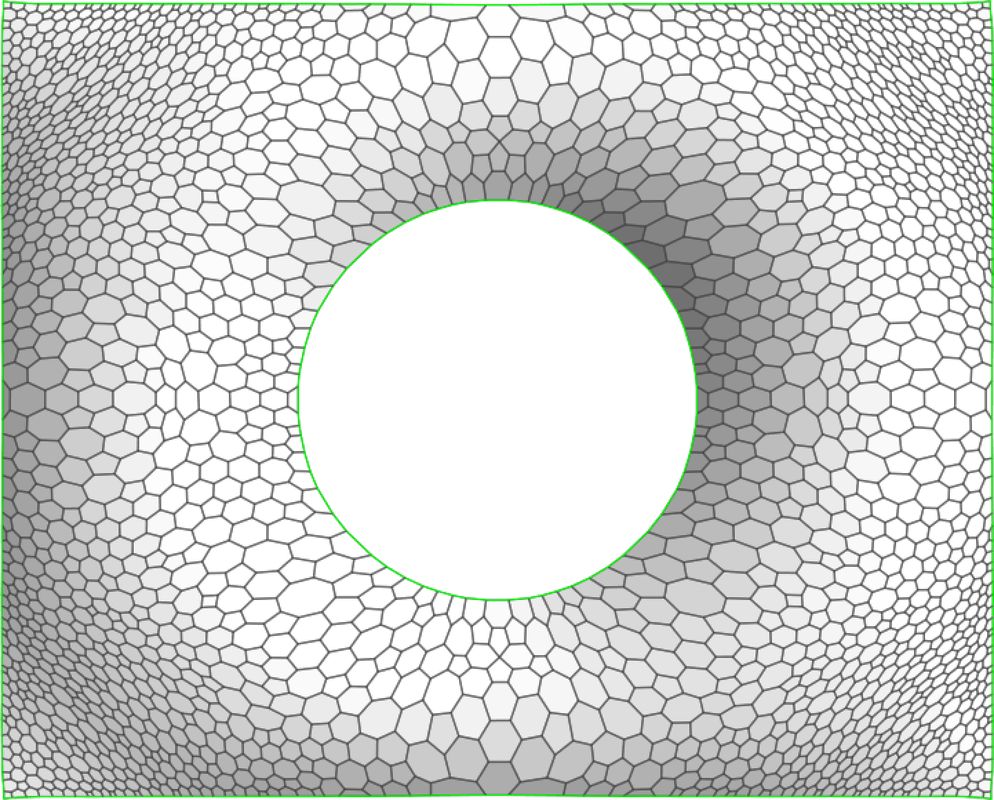}\\
	\cite{Tang:2014}&Anisotropic Quad & Voronoi Grid-Shell\\
	\includegraphics[height=0.25\columnwidth]{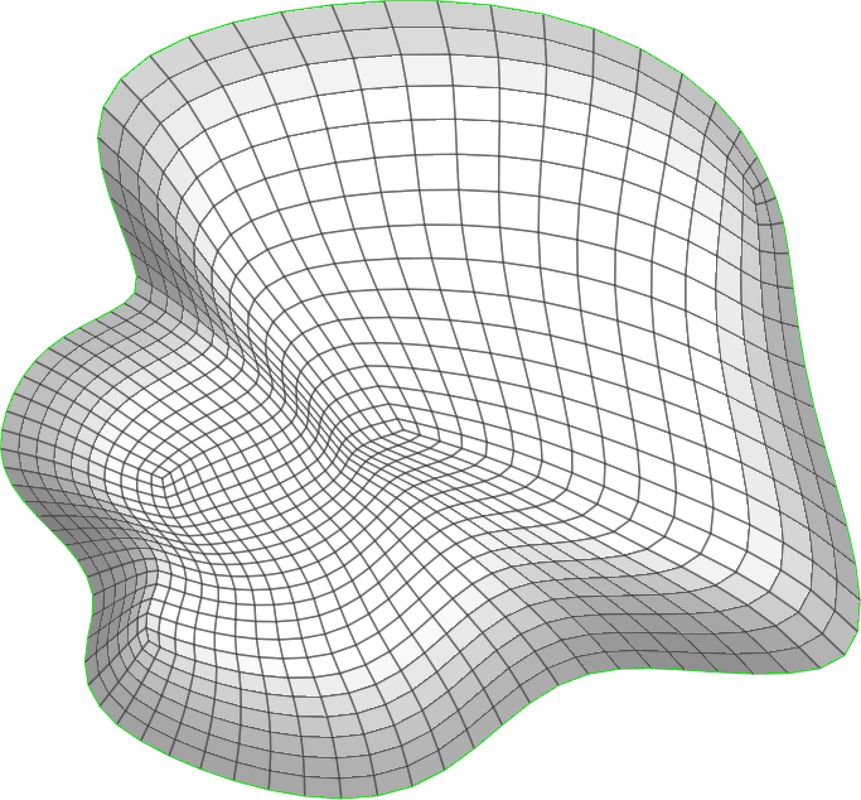}&
	\includegraphics[height=0.25\columnwidth]{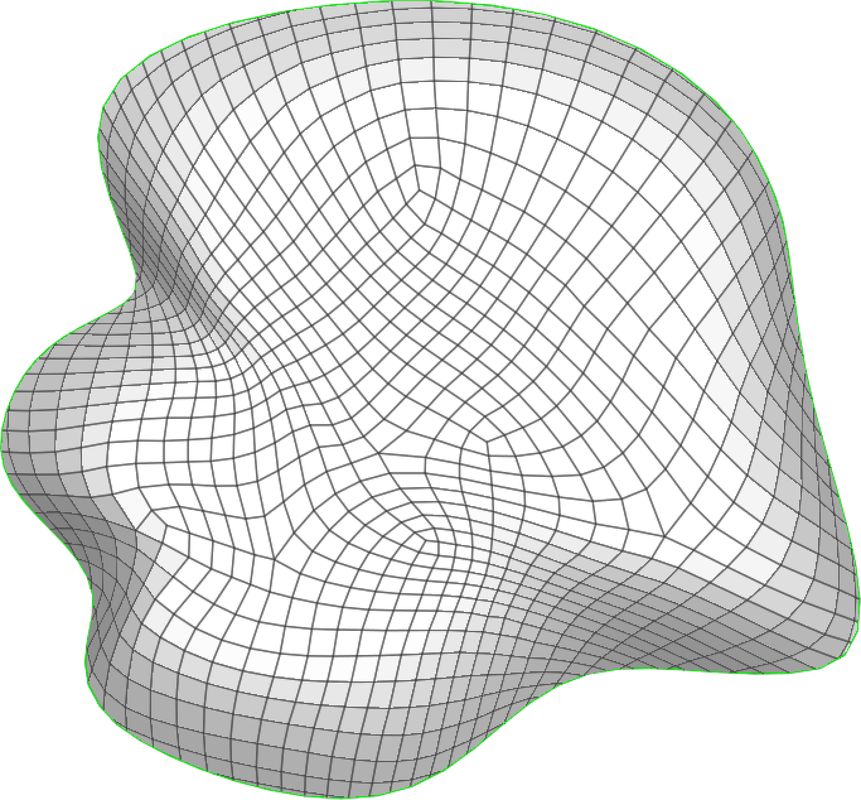}&
	\includegraphics[height=0.25\columnwidth]{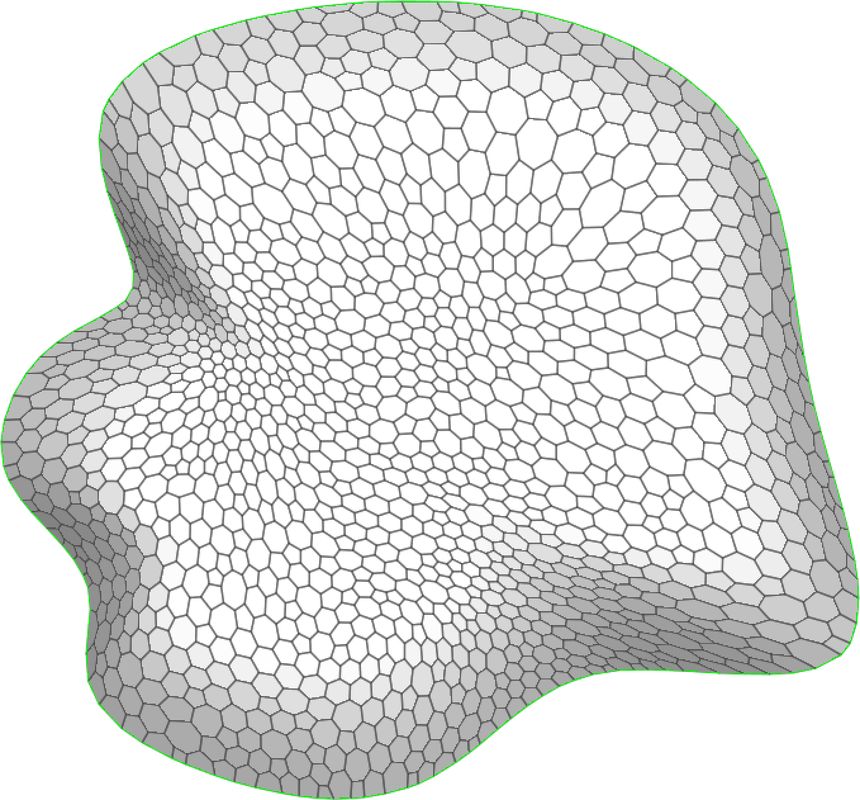}\\
	\cite{Tang:2014}&Anisotropic Quad & Voronoi Grid-Shell\\
	\includegraphics[height=0.25\columnwidth]{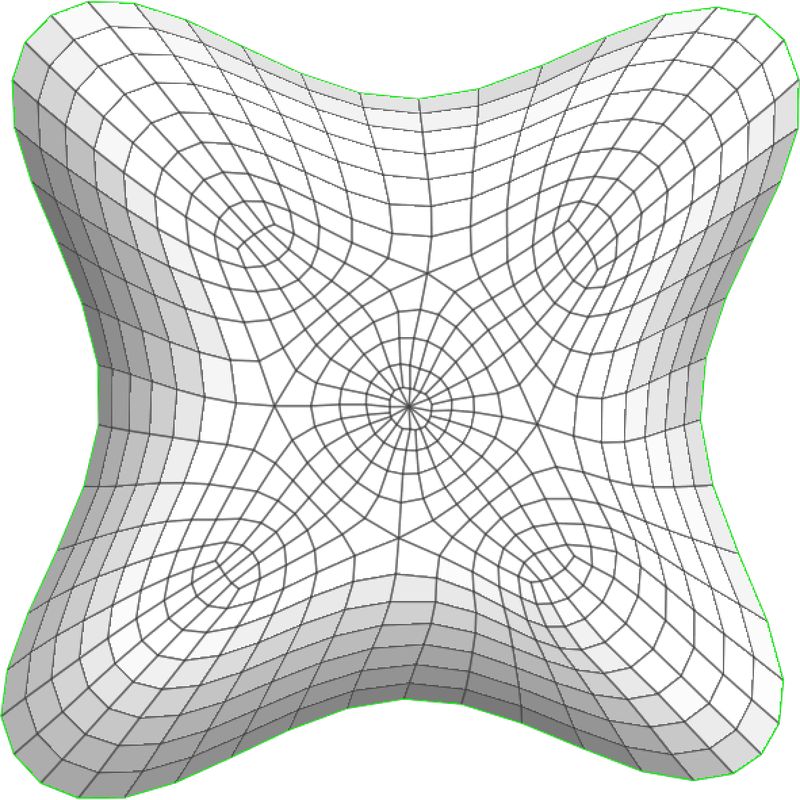}&
	\includegraphics[height=0.25\columnwidth]{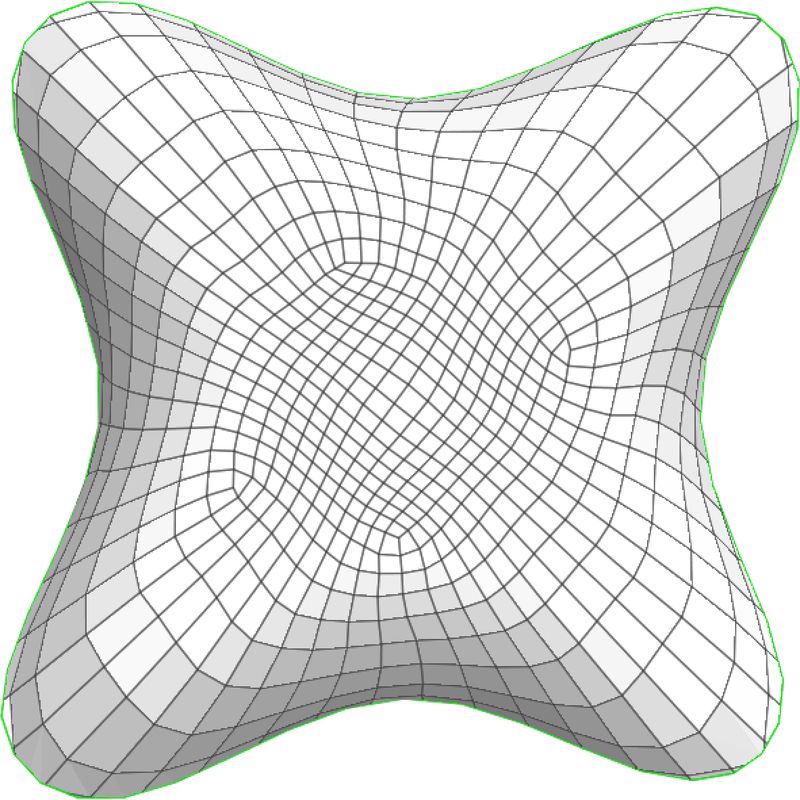}&
	\includegraphics[height=0.25\columnwidth]{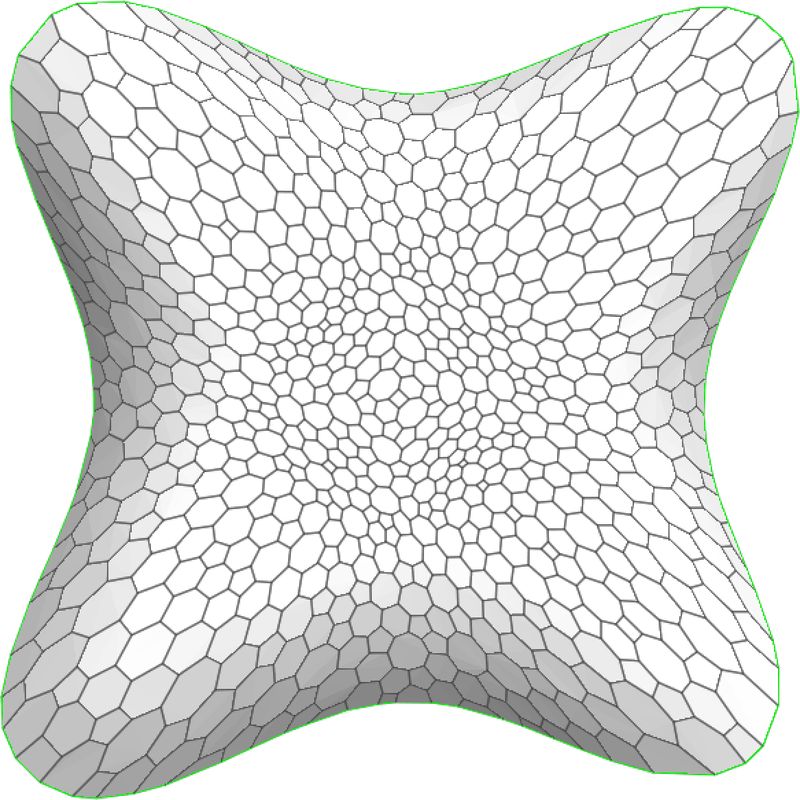}\\
	\cite{Vouga:2012}&Anisotropic Quad & Voronoi Grid-Shell\\
	\includegraphics[height=0.25\columnwidth]{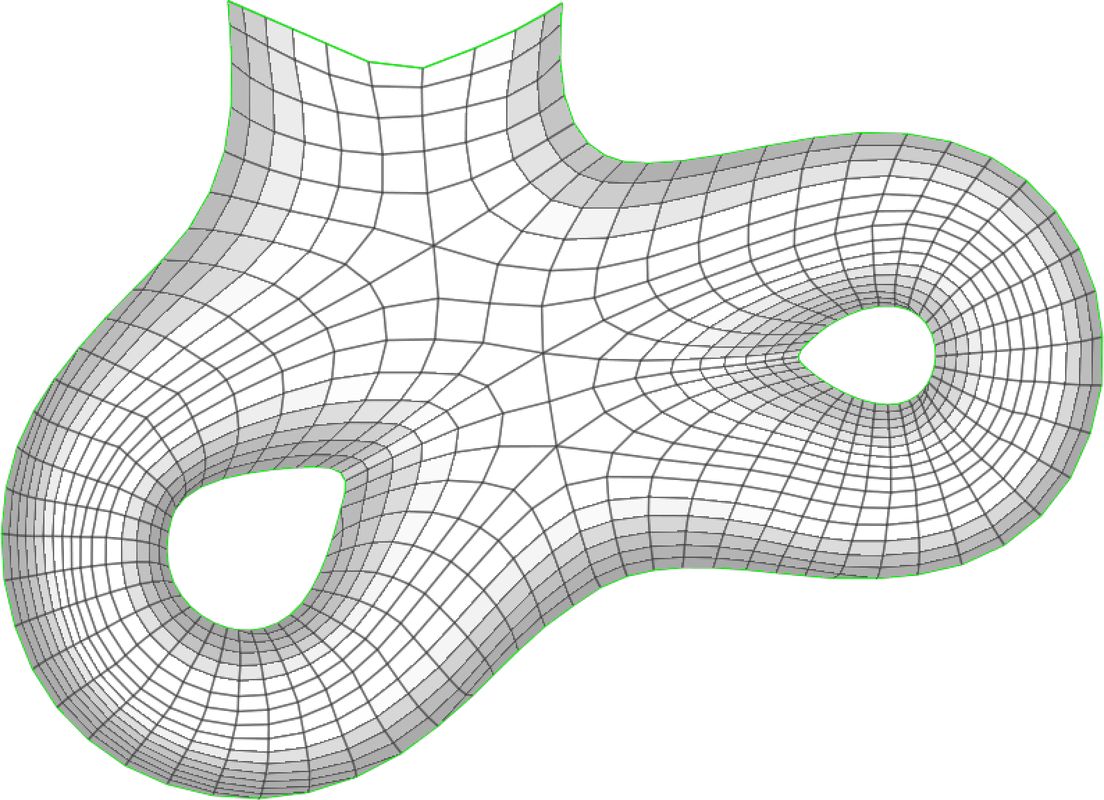}&
	\includegraphics[height=0.25\columnwidth]{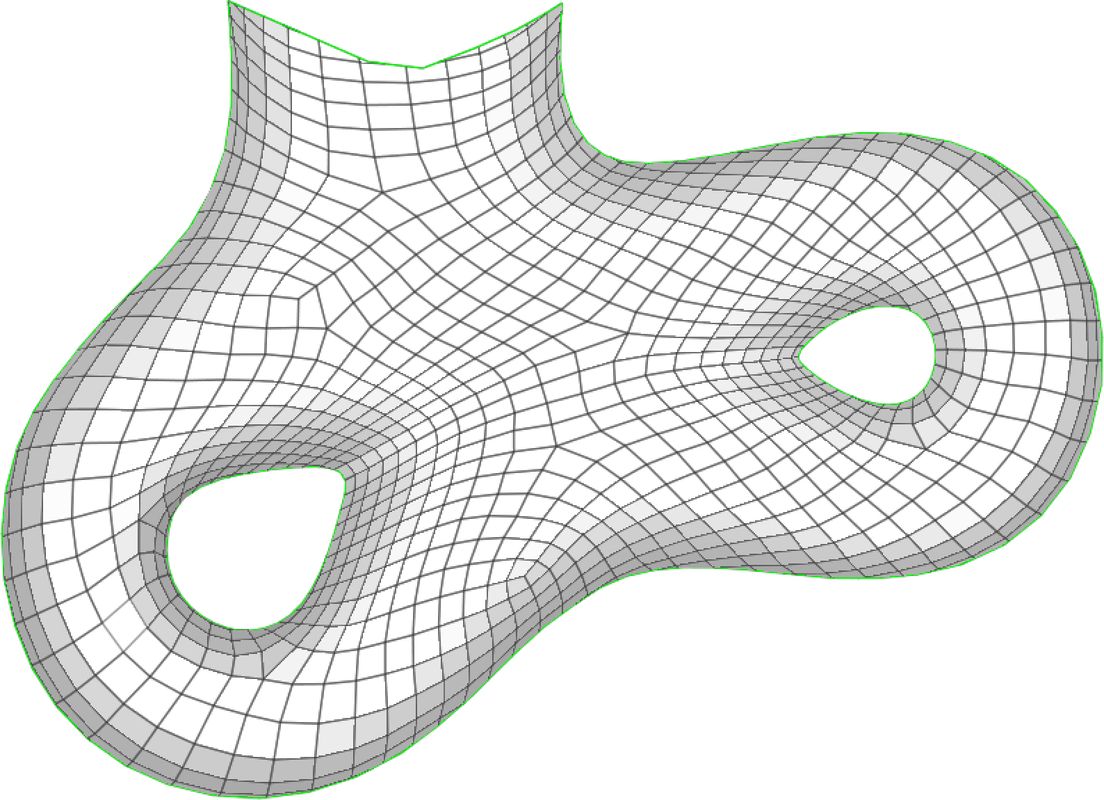}&
	\includegraphics[height=0.25\columnwidth]{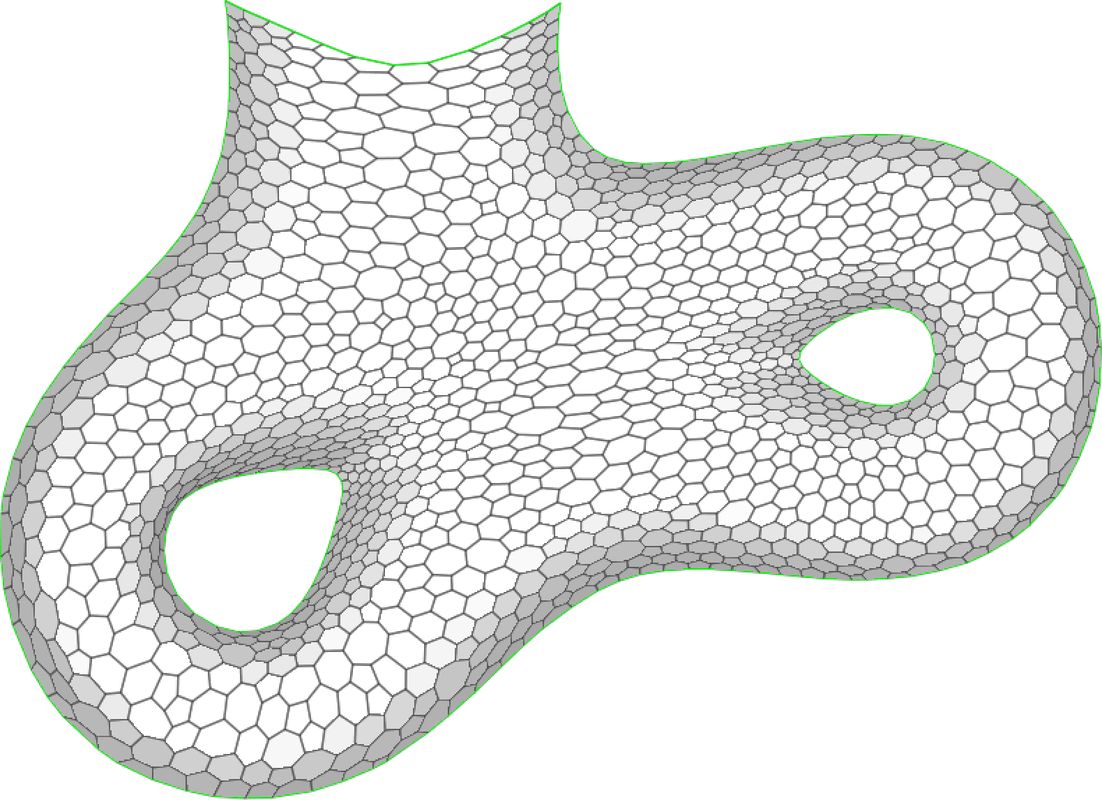}\\
	\cite{Vouga:2012}&Anisotropic Quad & Voronoi Grid-Shell\\
\end{tabular} 
	\caption{Comparison with quad meshes, top views. From the top: British Quad, Botanic, Lilium, Aquadom.} %The last model shows the original model for the Neum\"{u}nster  gridshell. }
	\label{fig:VsQuads}
\end{figure}
%===================================

\subsection{Comparison with Quadrilateral meshes}
We compared our grid-shells with some quadrilateral meshes obtained with  \cite{Tang:2014,Vouga:2012}. 
%To compare the structures, the displacement value is calculated at the maximum load supported by the quadrilateral meshing. 
As for the previous experiments, we set our parameters to match the total length of edges of the structures we compare with.
In order to evaluate how much benefit comes from the Voronoi approach, and how much from allowing for anisotropic and non-uniform meshing, we have also computed anisotropic quadrilateral meshes guided by the same stress tensor $\Psi$ that we use for our Voronoi structures. %, with the same $D$ and $A$ parameters. 
To this aim,  we have used the quadrangulation method in \cite{Panozzo:2014} by taking in input $\Psi$ (rescaled with the same $D$ and $A$ parameters we use for the ACVT) as guiding frame field.
The experiments are summarized in Table \ref{tab:datasets} and the related meshes are shown in Figure \ref{fig:VsQuads}. 
Our Voronoi grid-shells always achieve better performances, in terms of both buckling and displacement, than isotropic quad meshes obtained with state-of-the-art methods.
Voronoi grid-shells have also either better or comparable performances  with respect to our anisotropic quad meshes, with the only exception of the Botanic dataset, where the anisotropic quad mesh achieves a smaller displacement.
This suggests that both the Voronoi approach and the non-uniform anisotropic meshing play a role in improving performance.

\begin{figure}[htcb]
\centering
\begin{tabular}{ @{}c@{}c@{}c@{}c@{}} 
\includegraphics[width=0.5\columnwidth]{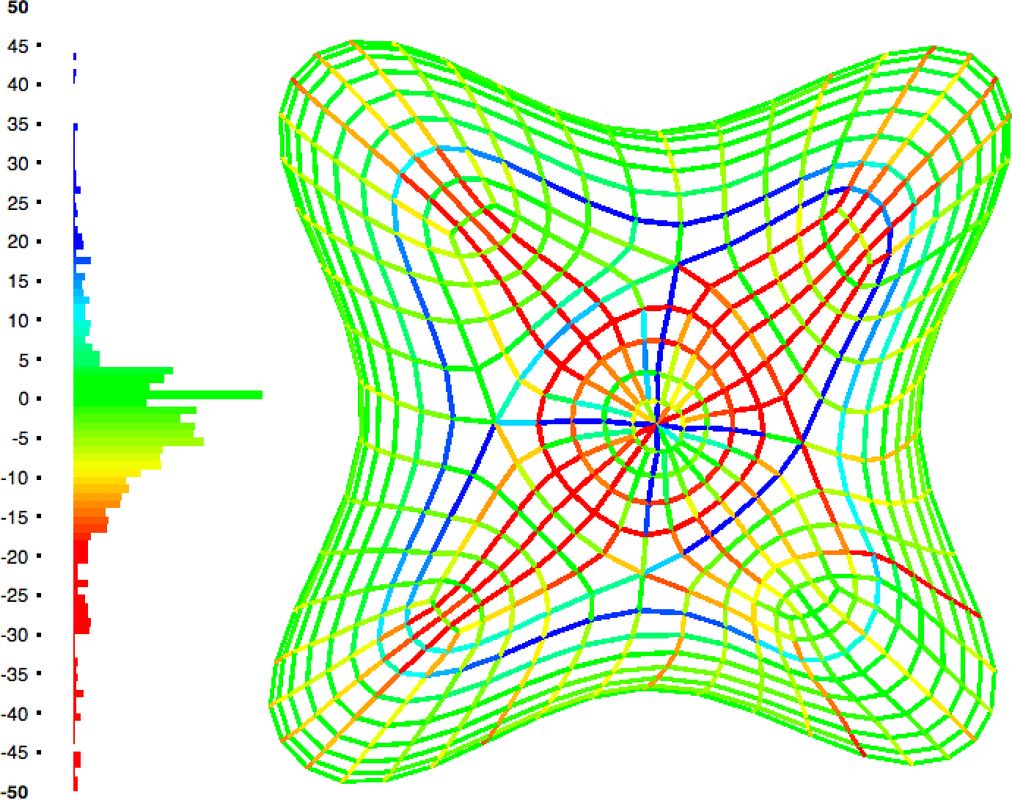}&
\includegraphics[width=0.5\columnwidth]{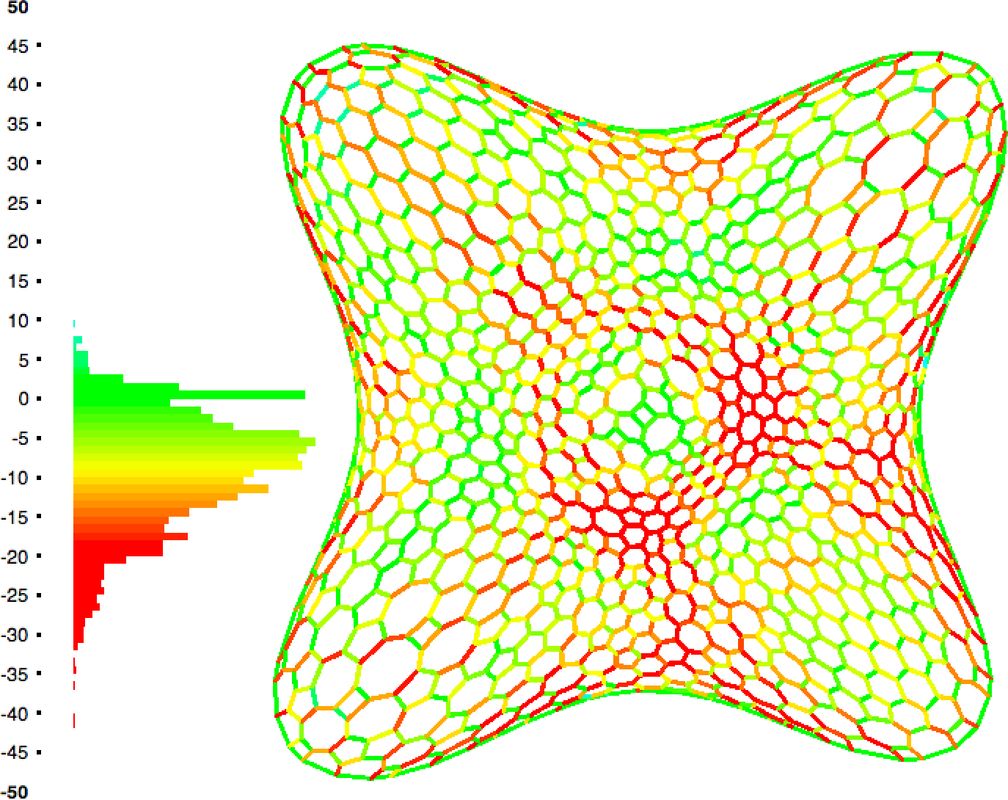}\\
~\\
\includegraphics[width=0.5\columnwidth]{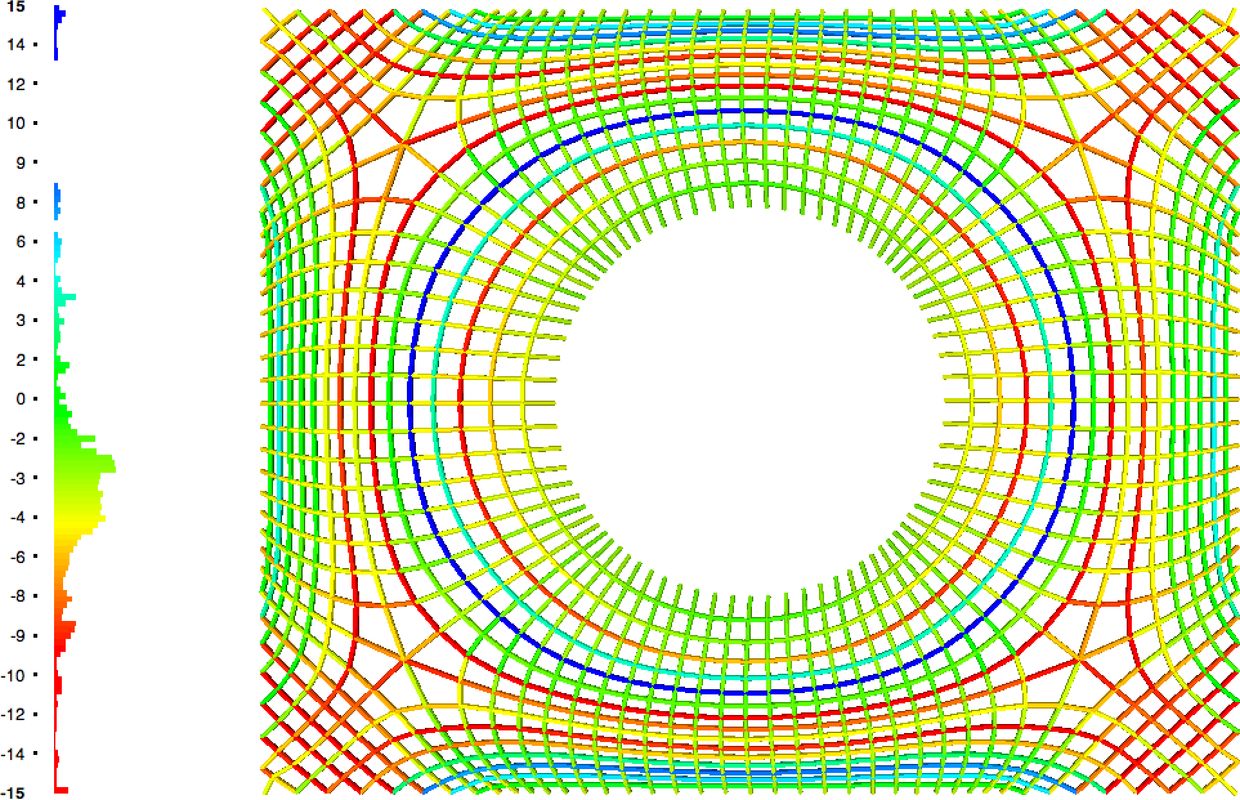}&
\includegraphics[width=0.5\columnwidth]{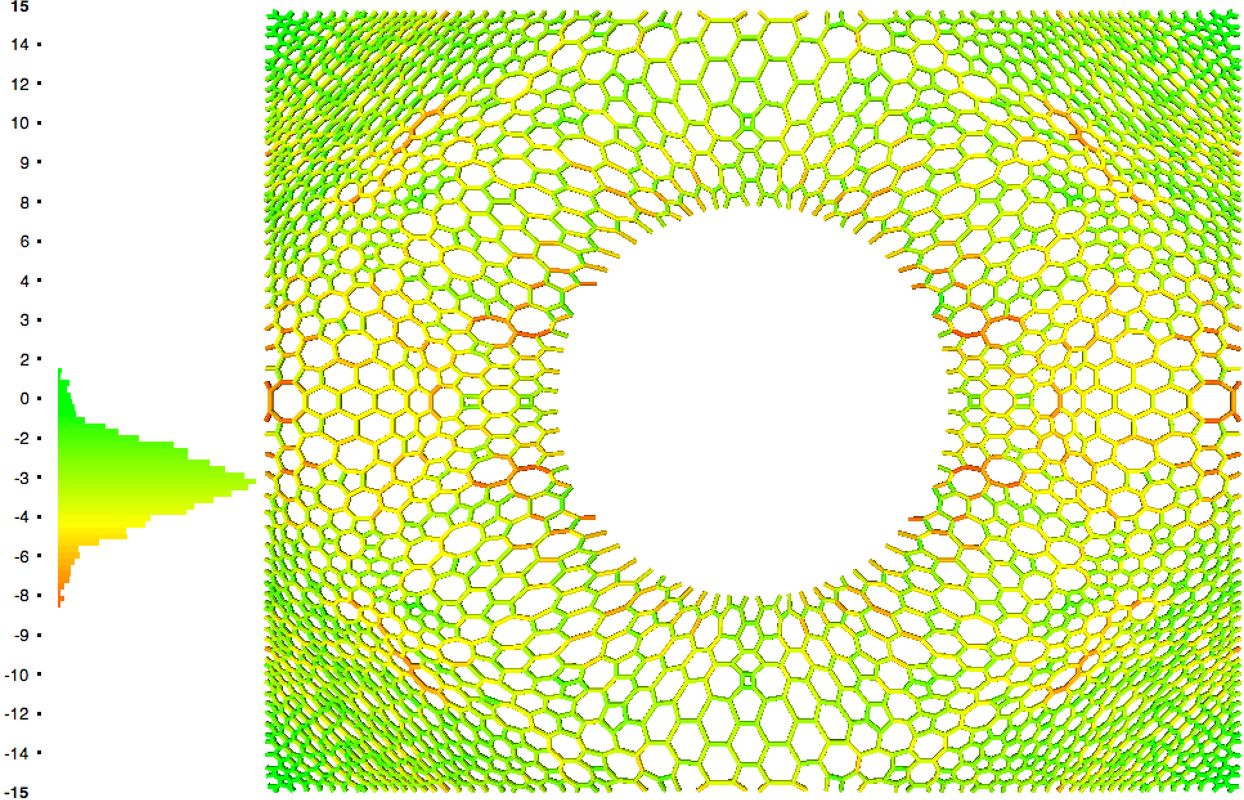}\\
\end{tabular} 
	\caption{Comparisons of distribution of AxialForces on the Lilium (top) and British Quad (bottom) datasets. Red corresponds to compression, blue corresponds to traction.}
	\label{fig:AxialForces}
\end{figure}

Figure \ref{fig:AxialForces} shows the effect of tessellation on the structural behavior of the grid-shell. 
%As you can see the Voronoi polygonal pattern displays a much more tridimensional load-bearing behavior than its quadrilateral counterpart.
In the Lilium dataset,  the forces flow from the top  to the restraints along the red paths of structural elements: 
in our model, such paths are better distributed, thus reducing the elastic strain energy W, as well as the maximal displacement. 
In the British Quad dataset, almost all the beams of our model undergo the same axial force, whereas in the quad model there is a strong variance of axial forces, including compressions (red) and traction (blue). 
%Therefore our Voronoi scheme allows an optimal usage of the material.
%===================================

%===================================

%===================================

\begin{figure}[htcb]
\centering
\begin{tabular}{ @{}c c@{}} 
\includegraphics[width=0.50\columnwidth]{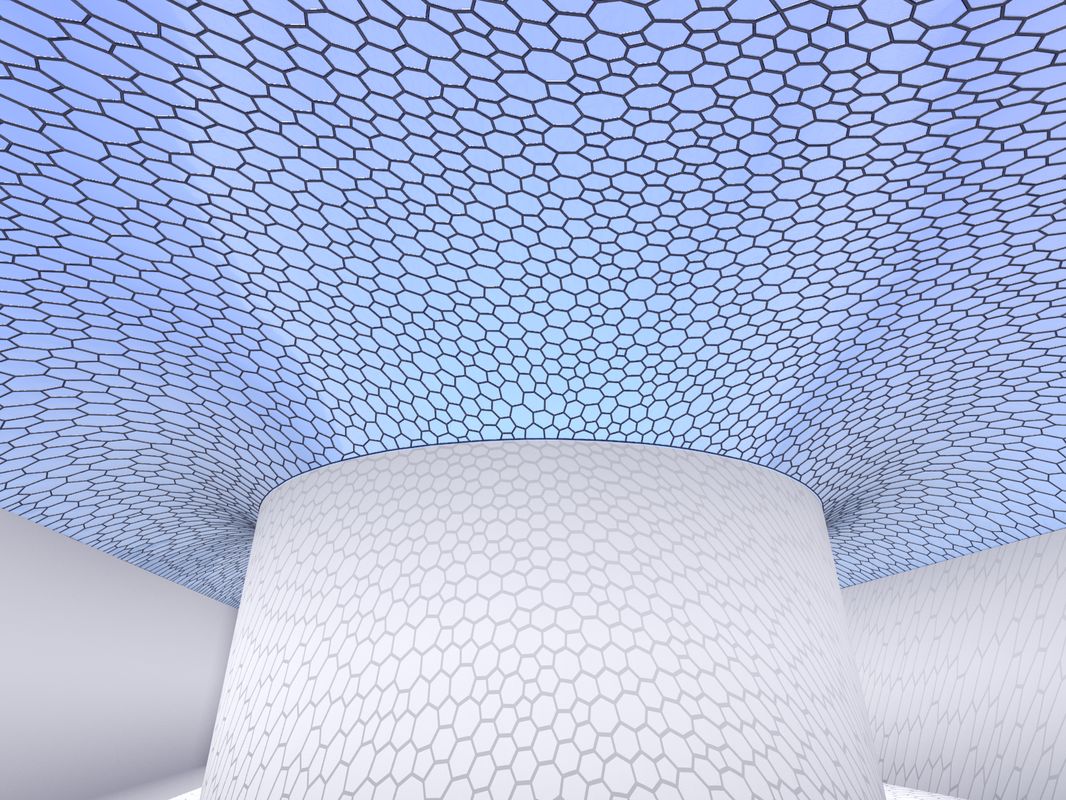}&
\includegraphics[width=0.50\columnwidth]{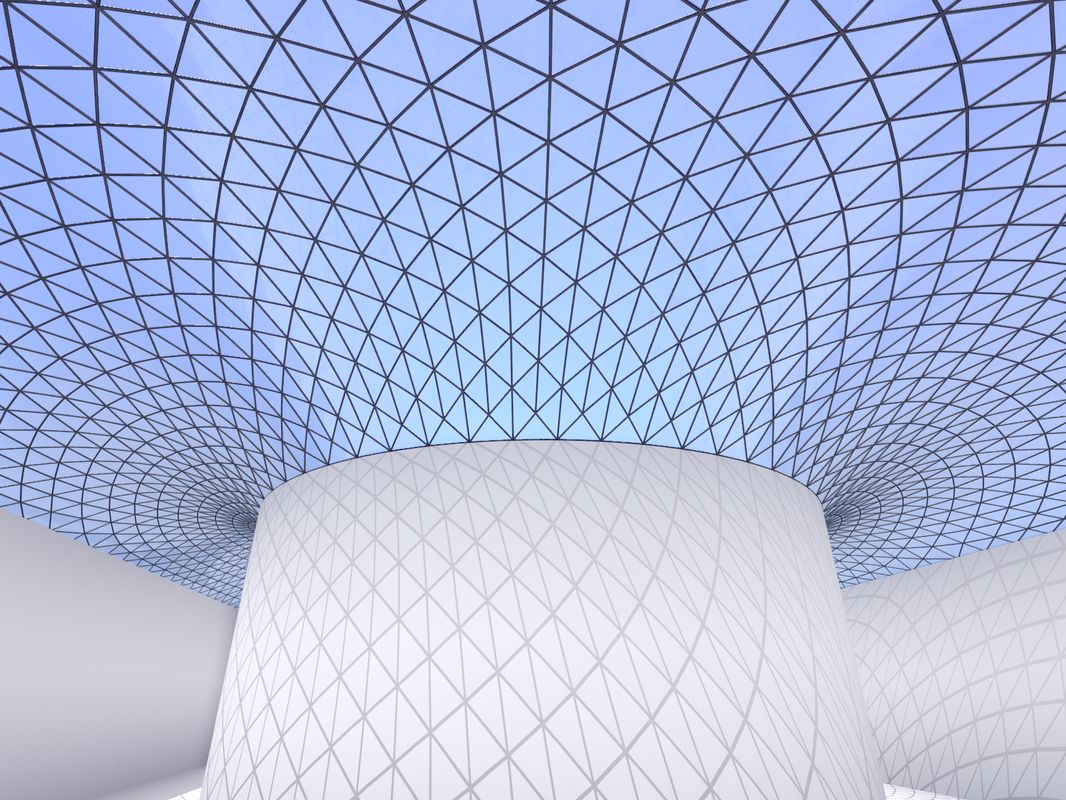}\\
Voronoi Grid-Shell & Original \\
$\lambda=6.82$ $\delta=15.09$&
$\lambda=9.62$ $\delta=2.6$\\
\end{tabular} 
	\caption{Comparison of our meshing of the British Museum versus the original triangulated mesh.}
	\label{fig:RenderingBritishTris}
\end{figure}

%===================================

\subsection{Comparison with Triangle meshes}
We also compared our structures with real examples of triangulated grid-shells. 
In Figure \ref{fig:RenderingBritishTris}, we show a comparison with the original meshing of the British Museum coverage (dataset British Tri, which is different from the British Quad surface considered in the previous experiment).
Related parameters can be found in Table \ref{tab:datasets}.
%and the ones that is has been used to built the Neum\"{u}nster Abbey. 
As expected, the triangular mesh has a better static behavior than our structure. 
The difference in terms of robustness is not dramatic, while the triangle-based grid-shell achieves a much better performance in terms of maximum displacement.
%though, and even the maximum displacement of our model, which is much larger than that of the super-rigid triangular mesh, it is about 15mm for a structure that spans over XXX meters, thus acceptable for practical fabrication purposes. 
%This is behavior is expectable as it is well know in structural engineering theta triangle mesh is the most robust structure. This is due to the fact that every possible deformation induces a compression or extension along one of the pipes.\\ 
%In particular, in Neum\"{u}nster  example shown by figure \ref{fig:RenderingTuning} we got a buckling factor of ? and ? for the triangular mesh. The comparison with the British museum is shown by figure ?, even in this case we got a buckling factor of  ? and ? for the triangulated version (see figure \ref{fig:RenderingBritishTri} ). \\ 

%\subsection{Aesthetics and light transport}
By considering this experiment, one may be tempted to deduce that architects should always rely on triangular grid-shell structures. 
However, triangular meshes are considered obsolete nowadays by architects both from an aesthetic and from a manufacturing point of view, while our Voronoi meshes offer an innovative design.
More generally, hex-dominant structures have several manufacturing advantages:
due to the lower valence of nodes, the joints are simpler to manufacture and assemble;
besides, it is possible, with further geometric optimization that slightly perturbs the original geometry, to obtain torsion-free structures \cite{Pottmann:2014}.

Moreover our patterns have a better perimeter/area ratio, therefore the average size of voronoi panels is significantly lower than the one of triangular meshes for the same total length of beams.
\begin{wrapfigure}{r}{0.35\columnwidth}
%  \begin{center}
  \vspace{-12pt}
  \hspace{-20pt}
    \includegraphics[width=0.40\columnwidth]{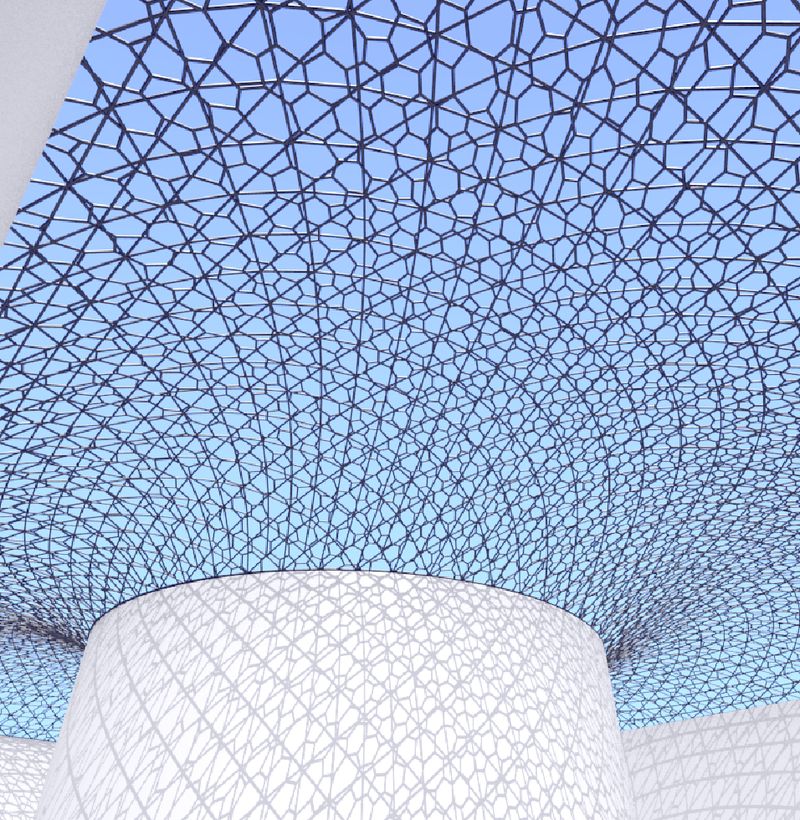}
    \vspace{-8pt}
%  \end{center}
\end{wrapfigure}
This can be clearly seen in the inset, where the two structures shown in Figure \ref{fig:RenderingBritishTris} are superimposed. 
Currently, one of the factors affecting the costs of the shell-grids is the size/radius of the used panels: the larger the size the higher the costs; our structures use  smaller panels for the same overall length, allowing for possible economic savings. 
%moreover, as shown by figure ??? hexagonal meshes allow for more light to seep through the envelope.

%begin{figure}[htcb]
%\centering
%\includegraphics[width=0.48\columnwidth]{./images/blender/british_tris_mix}
%\end{tabular} 
%	\caption{Comparison of our meshing of the British Museum versus the original triangulated mesh.}
%	\label{fig:RenderingBritishTris}
%\end{figure}

%\begin{wrapfigure}{r}{0.4\columnwidth}
%  \begin{center}
%    \includegraphics[width=0.48\columnwidth]{./images/blender/british_tris_mix}
%  \end{center}
%\end{wrapfigure}

%\begin{figure*}[htcb]
%\centering
%\begin{tabular}{ @{}c@{}c@{}}
%	\begin{tabular}{ @{}c@{}c@{}c@{}} 
%		\includegraphics[height=0.44\columnwidth]{./images/pictures/assembling.jpg}&
%		\includegraphics[height=0.44\columnwidth]{./images/pictures/final0.jpg}&
%		\includegraphics[height=0.44\columnwidth]{./images/pictures/final1.jpg}\\
%	\end{tabular} &
%	\begin{tabular}{ @{}c@{}} 
%		\includegraphics[height=0.214\columnwidth]{./images/pictures/close0.jpg}\\
%		\includegraphics[height=0.214\columnwidth]{./images/pictures/close1.jpg}\\
%	\end{tabular} 
%	\end{tabular} 
%	\caption{Fabricating a model. }
%	\label{fig:Fabrication}
%\end{figure*}

\begin{figure}[htcb]
\centering
\begin{tabular}{ @{}c@{}c@{}}
	\begin{tabular}{ @{}c@{}} 
		\includegraphics[height=0.5\columnwidth]{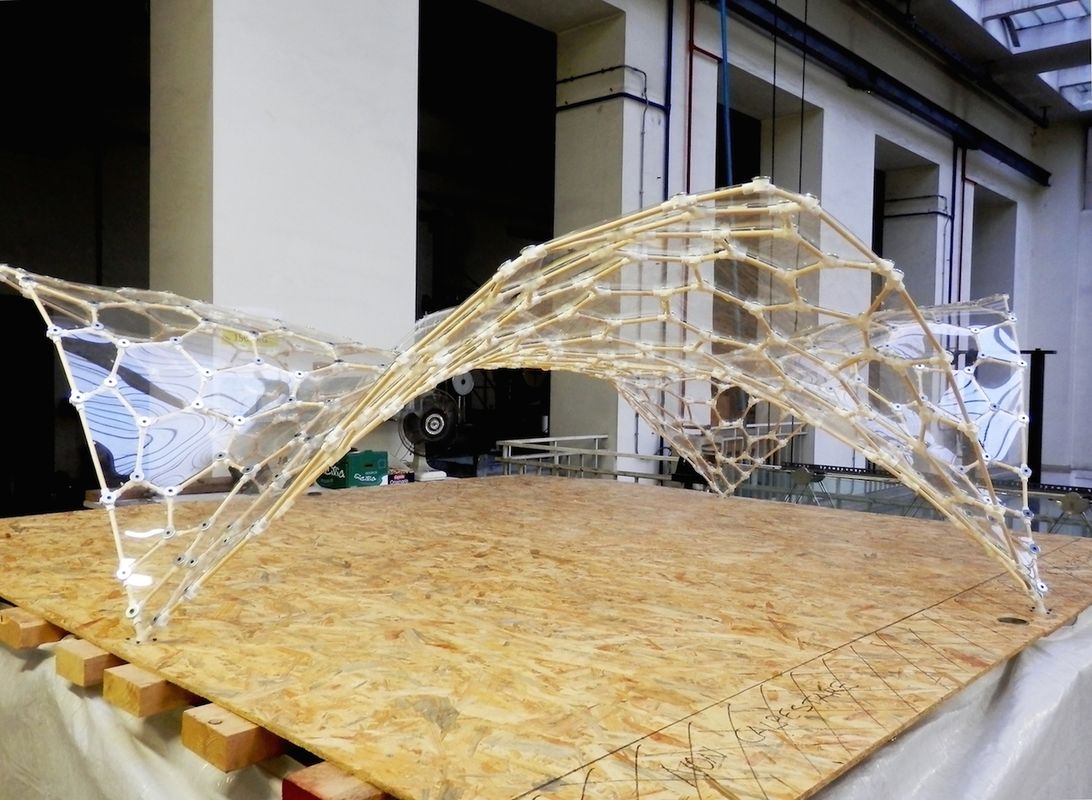}
	\end{tabular} &
	\begin{tabular}{ @{}c@{}} 
		\includegraphics[height=0.24\columnwidth]{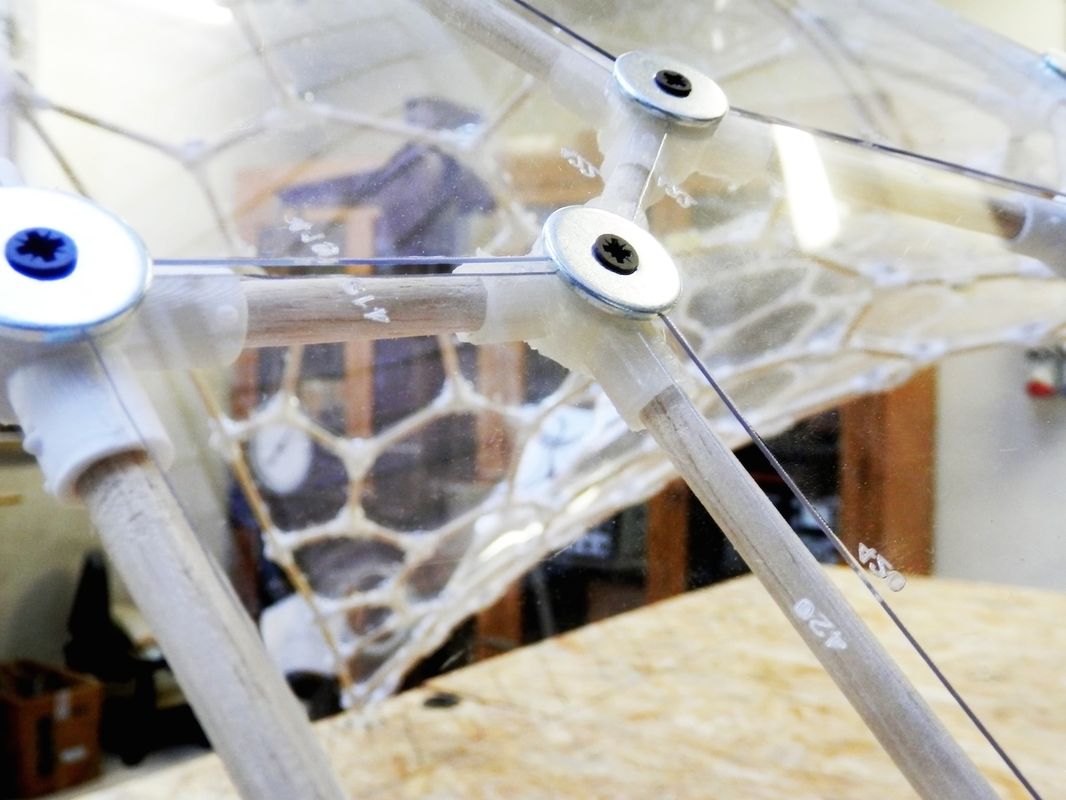}\\
		\includegraphics[height=0.24\columnwidth]{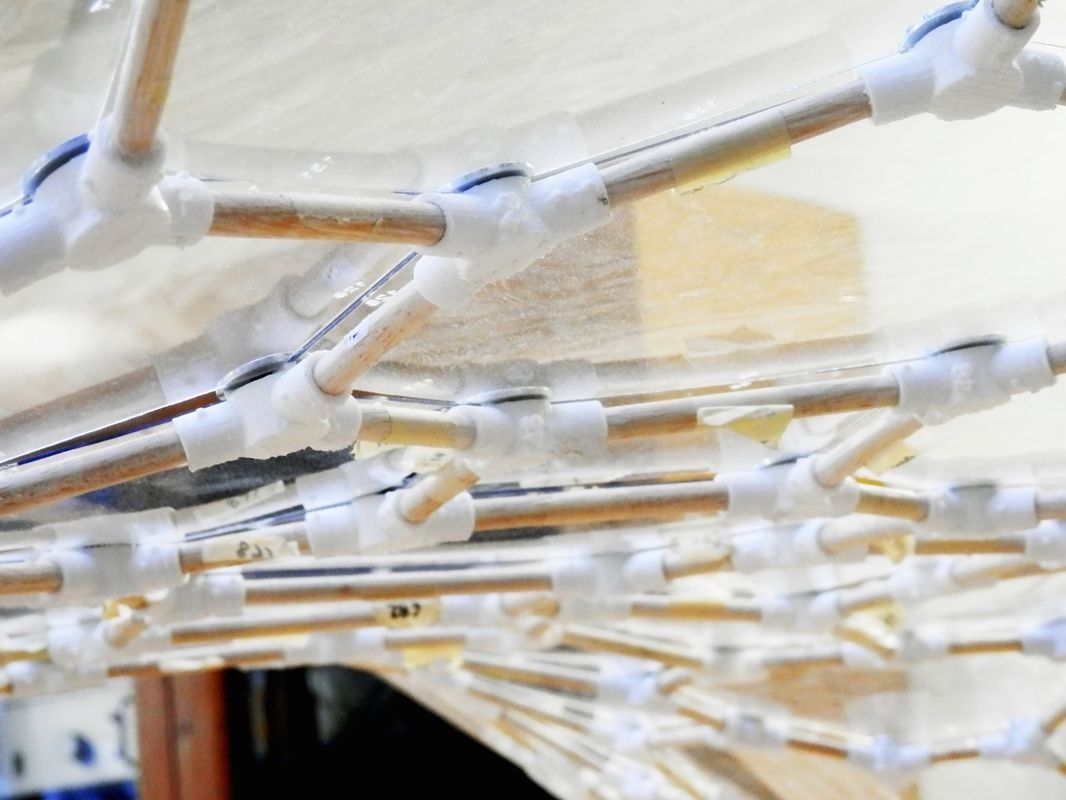}
	\end{tabular} 
\end{tabular} 
\caption{Fabricating a model of the Shell surface. }
\label{fig:Fabrication}
\end{figure}

\subsection{Physical replica} %roduction and assembly}

We have fabricated a reduced scale model of the Shell structure composed of 465 joints, 697 beams and 462 panels.
The side of this reproduction is 2.4 meters. 
Each joint has been produced independently using a FDM printer; sticks of wood simulate beams; % (see figure \ref{fig:Fabrication}). 
axternal panels are made of PET (Polyethylene terephthalate) and they have been laser-cut.  
Each component of the structure has a physical label (3D printed on joints, carved by laser on panels or glued paper on sticks), to help
%Labels allow 
us following a map to build the structure. 
The panels have been fixed by screwing a flat washer at each joint.
Some images of the %fabricated and mounted 
replica are shown in Figure \ref{fig:Fabrication}. 

\begin{figure}[htcb]
\centering
\begin{tabular}{ @{}c@{}}
\includegraphics[width=0.8\columnwidth]{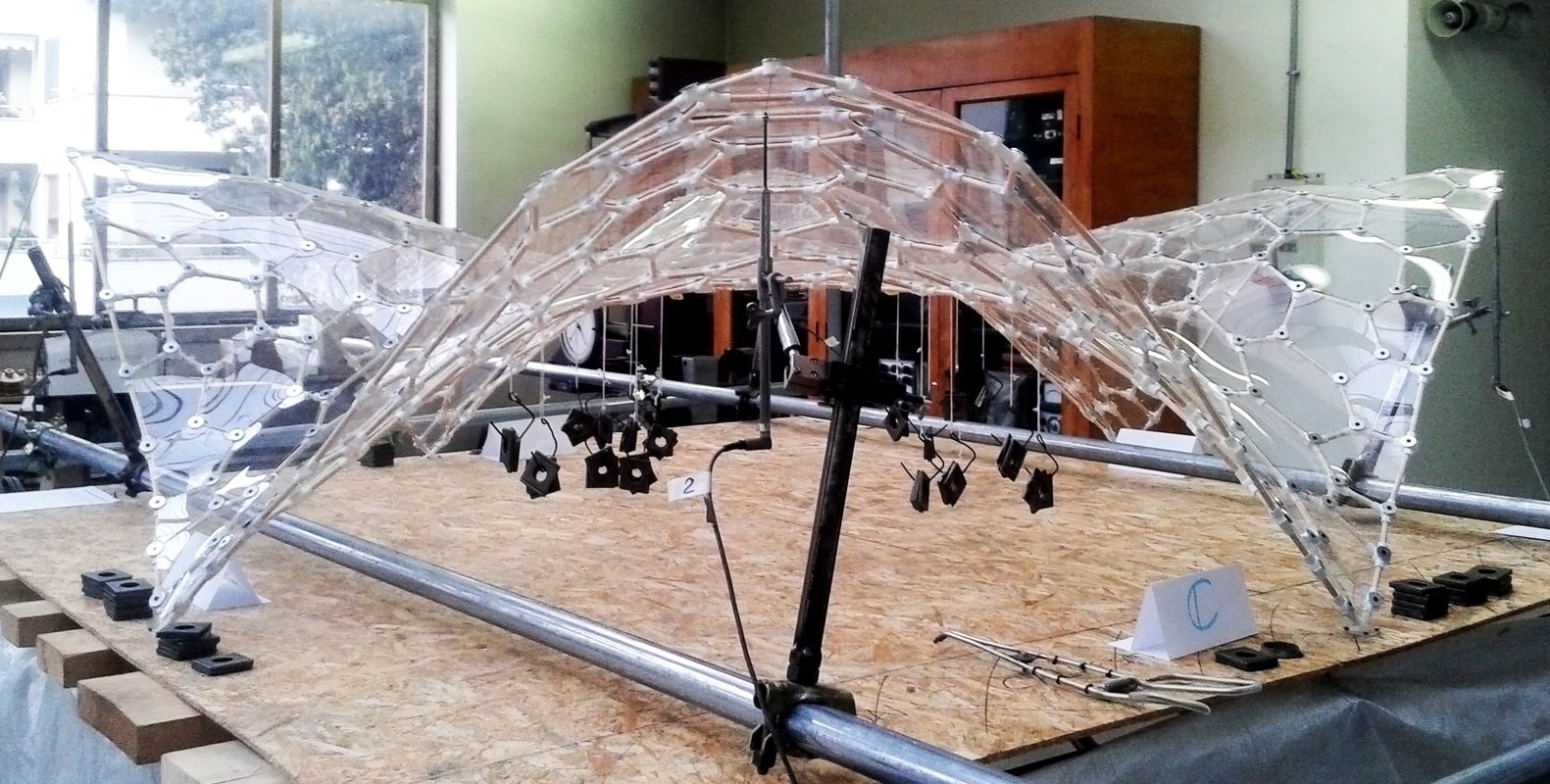}\\
\includegraphics[width=0.8\columnwidth]{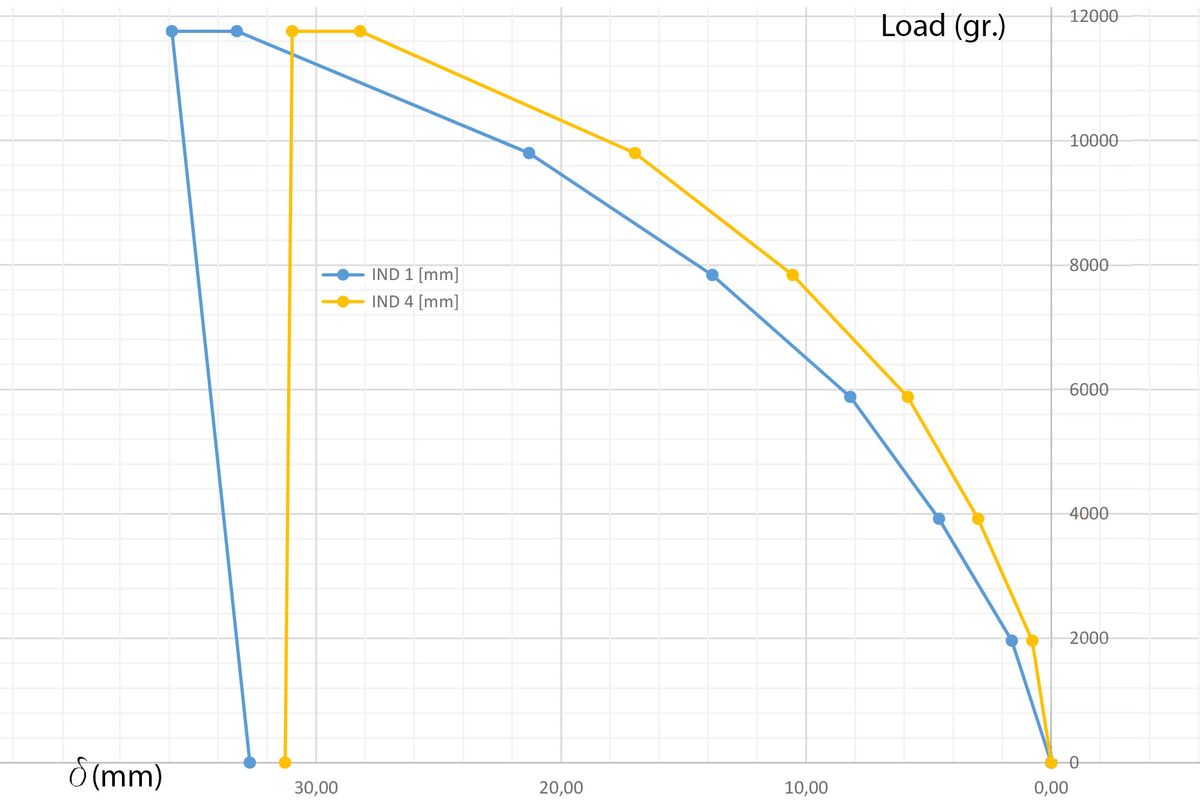}\\
\end{tabular}
	\caption{Top: Setup for load tests over the fabricated model; Bottom: displacement / external load plot. }
	\label{fig:LoadTest}
	
\end{figure}

We have performed load tests on the physical structure, by incrementally applying weights and measuring the displacement of the structure with a proper sensor
%This setup is illustrated in 
(see Figure \ref{fig:LoadTest}).  
We have have monitored the displacement of the corners of the structure while gradually incrementing the external load. 
The result of this experiment is shown in the graphs of Figure \ref{fig:LoadTest}. 
Obviously we have relied on different materials (wood and ABS, rather then steel), however the general trend of deformation is similar to the simulated mesh.%We are planning to further investigate this aspect in future work.

%\TODO{ "similar" non e' un po troppo generico? }

%============================== BIG FLOATS ===============================

%===================================

%===================================

\begin{figure*}[tcbp]
\centering
\begin{tabular}{ @{}c@{}c@{}c@{}} 
\includegraphics[width=0.7\columnwidth]{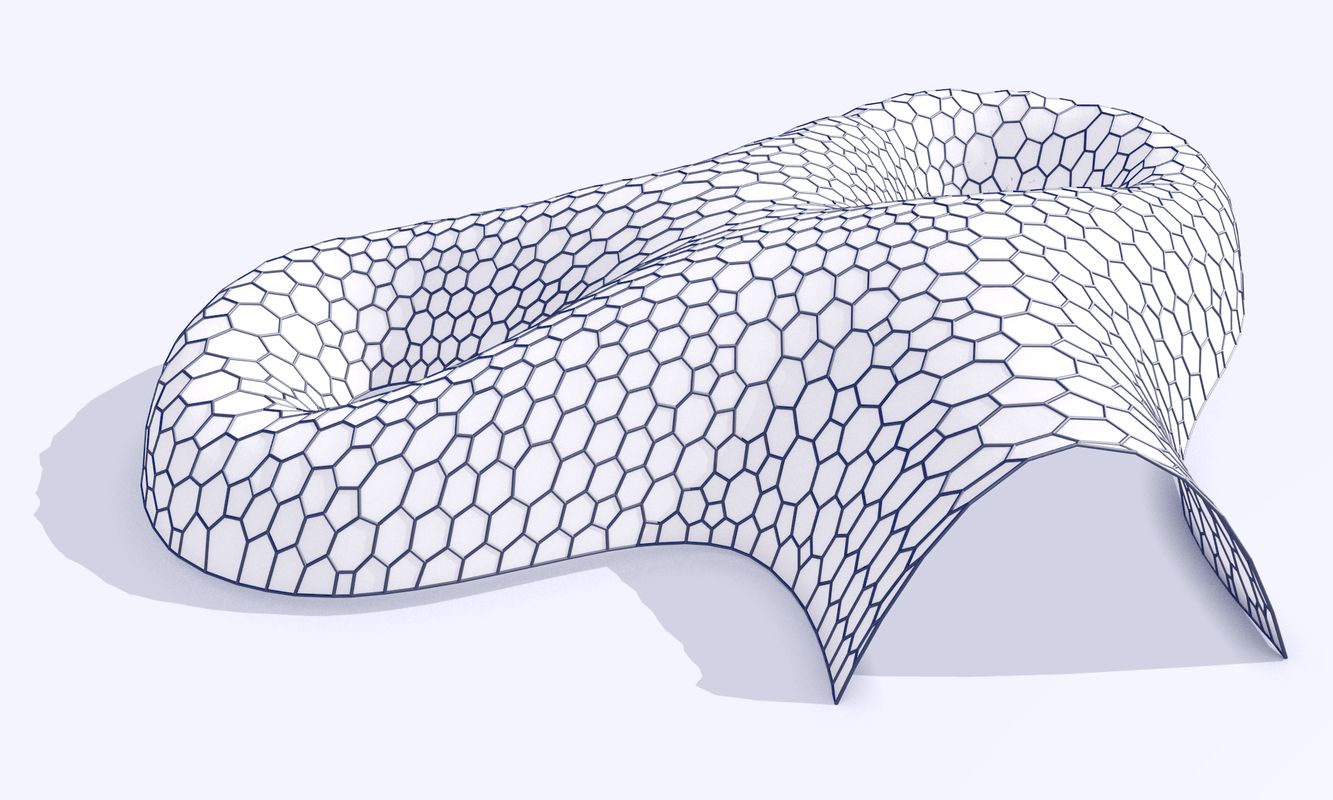}&
\includegraphics[width=0.7\columnwidth]{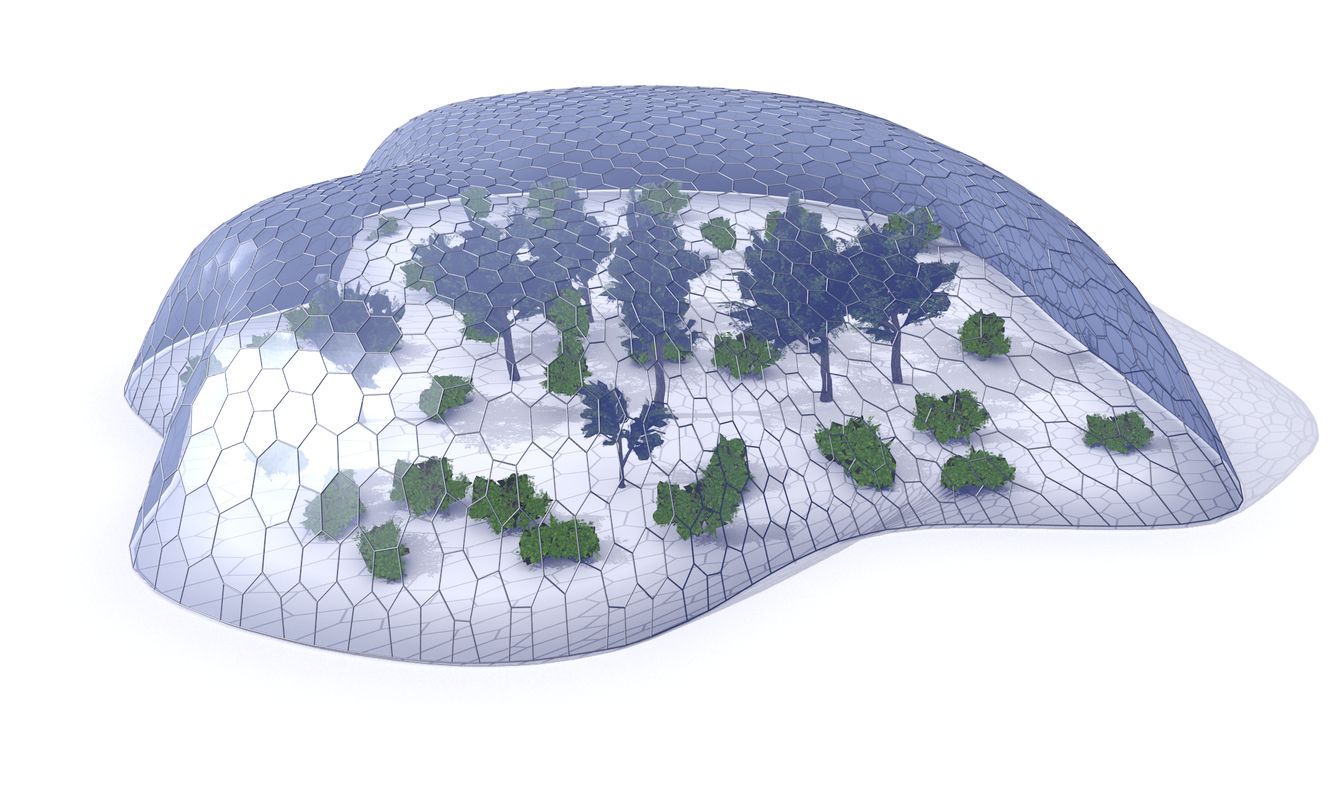}&
\includegraphics[width=0.7\columnwidth]{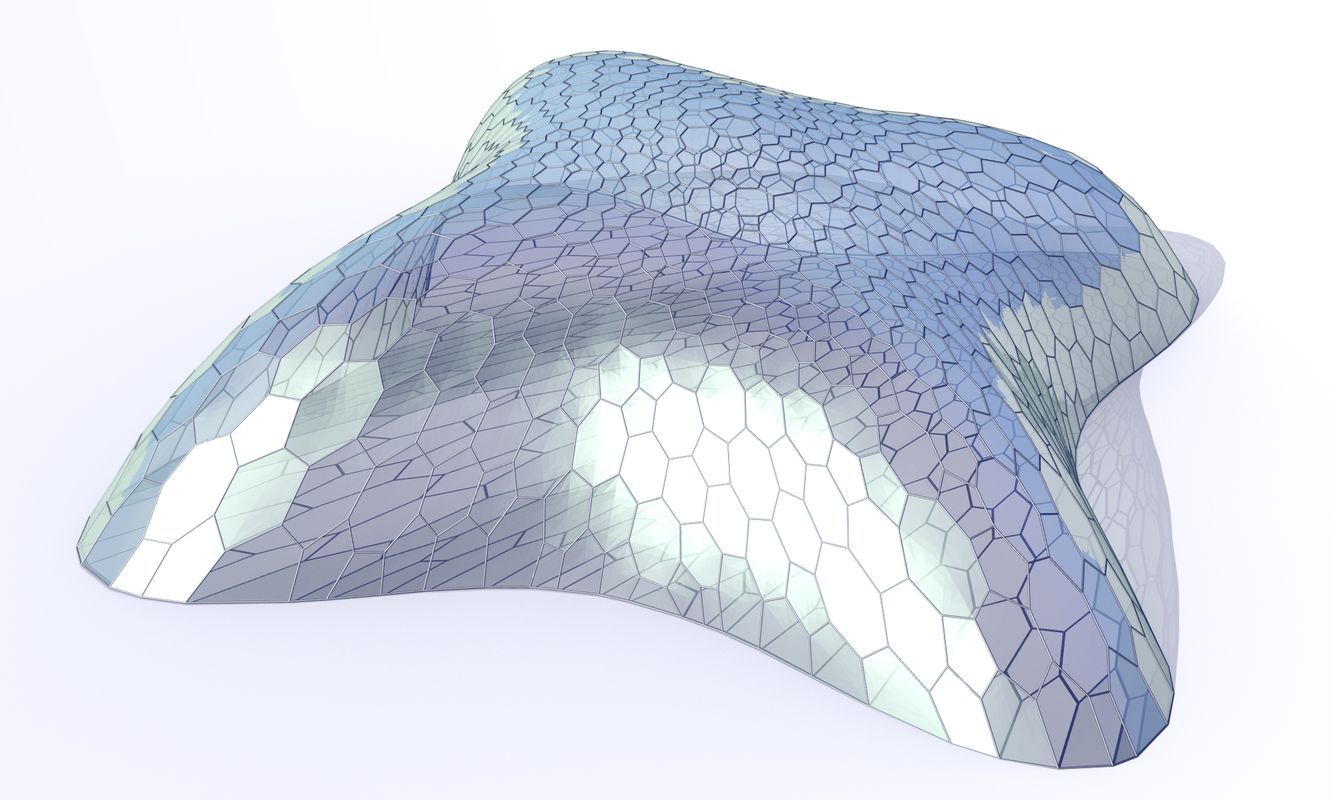}\\
\end{tabular} 
	\caption{Some of the models used for comparison with quadrilateral meshing. From the top: Aquadom, Botanic and Lilium.} 
	%The last model shows the original model for the Neum\"{u}nster  gridshell. }
	\label{fig:RenderingComparison}
\end{figure*}

\section{Concluding remarks}\label{sec:conc}
We have presented a practical and physically sound framework for the generation of grid-shell structures whose topology is based on optimized Anisotropic Centroidal Voronoi Tessellations. 
%The process is driven by a initial static analysis of the geometric shape of the surface. 

We use the tensor field resulting from FEM stress analysis on the input surface to induce an anisotropic non-Euclidean metric over it. 
Then we compute an Anisotropic Centroidal Voronoi Tessellation under the same metric. 
The resulting mesh is hex-dominant and made of cells with variable density, depending on the amount of stress, and anisotropic shape, oriented along directions of maximum stress. 
This mesh is further optimized taking into account symmetry and regularity of cells to improve aesthetics.

We have tested the generated structures evaluating, by means of industrial standard non-linear analysis simulations, their behavior in terms of non-linear buckling multiplier and nodal displacement. 
We have built a reduced scale model and we have performed physical tests on it to verify the soundness of the behavior predicted by the simulation.
% with the real one. 
The result of our experiments demonstrate that our grid-shells achieve better static performances with respect to quad-based grid-shells, while offering an innovative and aesthetically pleasing look.

%\section*{Acknowledgements}
%Omitted for sake of anonymity. %messi per calcolo spazio
\bibliographystyle{acmsiggraph}
\bibliography{voronoi}
\end{document}